\documentclass[11pt,journal]{IEEEtran}

\usepackage{geometry}
\usepackage{sidecap}
\sidecaptionvpos{figure}{t}

\usepackage{amsfonts}
\usepackage{amsmath}
\usepackage{amssymb}
\usepackage{algorithm,algorithmic}
\usepackage{array}
\usepackage[caption=false,font=normalsize,labelfont=sf,textfont=sf]{subfig}
\usepackage{textcomp}
\usepackage{stfloats}
\usepackage{xspace}
\usepackage{caption}
\usepackage{tikz}
\usepackage{pifont}
\usepackage{inconsolata}

\usepackage{setspace}
\usepackage{mathtools}
\usepackage{amsthm}
\usepackage{url}
\usepackage{verbatim}
\usepackage{graphicx}
\usepackage{euscript}
\usepackage{cite,soul}
\usepackage{mathrsfs,xspace,multicol}
\usepackage[table]{xcolor}
\usepackage{tabularx}
\usepackage{booktabs}
\usepackage[normalem]{ulem}
\usepackage[inline, shortlabels]{enumitem}
\setlist{itemjoin ={,\enspace},itemjoin* = {, and\enspace}}
\usepackage{dutchcal}

 \usepackage[font=footnotesize]{caption}

\usepackage{hyperref}
\hypersetup{
colorlinks=true,
linkcolor=black,
filecolor=black,      
urlcolor=black,
citecolor =black,
}

\def\tk  									{t_k}
\def\psnr     							 {\mathsf{PSNR}\xspace}

\def\g                                      {g}
\def\f                                      {f}
\def\hf                                     {\widehat{f}}
\def\dB									    {dB}
\def\srf              					   {SRF\xspace}

\def\pr					                   {p}
\def\sr 								   {s_{K}}
\def\srr 								  {s_{\mathbf{r}}}
\def\tsr 								  {\widetilde{s}}
\def\gr 								  {g}
\def\grr 								 {g_{\mathbf{r}}}
\def\yr 								  {y}
\def\yrr 								 {y_{\mathbf{r}}}
\def\byr 								{\bar{y}}
\def\mbyr 							 {\bar{\mat{y}}}

\def\rrr 								  {r_{\mathbf{r}}}
\def\phir            					{\varphi}
\def\gf                                 {\phi}
\def\l										{\left(}
\def\r								       {\right)}
\def\iZ									 {\in \mathbb Z}
\def\iR								     {\in \mathbb R}

\def\Z									  {\mathbb Z}
\def\R									  {\mathbb R}
\def\C									  {\mathbb C}

\def\DE					                {\stackrel{\rm{def}}{=}}
\def\eg					                {\emph{e.g.\xspace}}
\def\ie					         	      {\emph{i.e.\xspace}}
\def\viz							     {\emph{viz.\xspace}}
\def\SR                 		        {{SRes}\xspace}
\def \dk           		   	 	         {\mathscr{D}_K}

\def\ttd								 {$3$D }
\def\gammar            			{\Gamma}
\def\gammarr            	   {\Gamma_{\mathbf{r}}}
\def\tgammar            		{\widetilde{\Gamma}}
\def\mgammar            	 {\boldsymbol{\Gamma}}
\def\mtgammar            	{\boldsymbol{\widetilde{\Gamma}}}
\def\taur            				   {\tau}
\def\taurr            				  {\tau_{\mathbf{r}}}
\def\ttaur            				  {\widetilde{\tau}}
\def\mtaur            				{\boldsymbol{\tau}}
\def\mttaur            			   {\boldsymbol{\widetilde{\tau}}}
\def\kernel							{\psi}
\def\fr              					  {\kernel}
\def\bfr              					{\bar{\kernel}}
\def\usf					   		   {USF\xspace}
\def\g               				 	 {g}
\def\HDR							{HDR\xspace}
\def\HDRes					     {HDRes\xspace}

\def\sritersis			 		  {\SR-IterSiS\xspace}
\def\USSR					      {FP-SR\xspace}
\def \ussr 							{US-SRes\xspace}
\def\ak 							   {\widehat{\kernel}_{i}}
\def\RMs		        		   {\mathsf{H}}
\def\ai								   {a_{i}}
\def\ma             				{\mat{a}}
\def\Ifr							   {I_{\fr}}
\def\Ofr							 {\Omega_{\fr}}
\def\taum					 	 {\tau_{m}}
\def\cm 					   		{c_{m}}
\def\mcm 					   	{\mat{c}}
\def\res          			   		{{\varepsilon}_{\gr}}
\def\bres          			   	  {\bar{\varepsilon}_{\gr}}
\def\btres          			 {\bar{\widetilde{\varepsilon}}_{\gr}}
\def\mbres          		   {\bar{\boldsymbol{\varepsilon}}_{\gr}}
\def\mbtres          		  {\bar{\boldsymbol{\widetilde{\varepsilon}}}_{\gr}}
\def\Mr								{M}

\def\interval				   {\mathcal{I}}
\def\SD						  	  {\mathcal{S}_{T}}
\def\hfr						   {\widehat{\kernel}}
\def\hgr						  {\widehat{\g}}
\def\htgr						 {\widehat{\widetilde{\g}}}
\def\bgr						  {\underline{\g}}
\def\tgr						   {{\widetilde{\g}}}
\def\mbgr					   {\underline{\mat{\g}}}
\def\mbtgr					  {\underline{\mat{\widetilde{\g}}}}
\def\hsr						  {\widehat{s}}
\def\htsr						 {\widehat{\widetilde{s}}}
\def\akl 						   {\widehat{\kernel}_{l}}
\def\uk							   {u_k}
\def\um							  {u_m}
\def\mf							  {\mat{f}}
		
\def\nm							 {n_{m}}
\def\mnm					  {\mat{n}}
\def\PO							{{\mathsf{P1}}}
\def\PT                			{{\mathsf{P2}}}

\def\bcoef				     {\mat{b}}		
\def\MSE					 {\mathsf{MSE}}
\def\cz							{c_{0}}
\def\Id							 {\mathbf{I}_d}
\def\rank					 {\mathsf{rank}}
\def\madc				  {{\scalefont{0.9}{$\mathscr{M}_\lambda$--\textsf{ADC}}}\xspace}
\def\L 							{L}
\def\coef					{b}

\def\zn						  {\zeta}

\newcommand{\norm}[1]           {\| {#1} \|}
\newcommand{\fig}[1]			    {Fig.~\ref{#1}}
\newcommand{\EQc}[1]		    {\stackrel{\eqref{#1}}{=}}
\newcommand{\id}[1]					{\mathbb{I}_{#1} }
\newcommand{\sqb}[1]		    {\left[ #1 \right]}
\newcommand{\Poly}[2]          {\RMs\rob{#2;#1}}

\newcommand\rob[1]				  {\l #1 \r}
\newcommand\secref[1]		   {Section \ref{#1}}
\newcommand\tabref[1]		    {Table \ref{#1}}
\newcommand\algref[1]			{Algorithm \ref{#1}}
\newcommand\thmref[1]		  { Theorem~\ref{#1}}
\newcommand\lemmaref[1]   {Lemma \ref{#1}}

\newcommand{\bpara}[1]		 {\smallskip \noindent {\bf #1}}

\newcommand{\transp}		  {{\top}} 
\newcommand{\Hermit}		 {^{\mkern-1.5mu\mathsf{H}}} 
\newcommand{\mat}[1]		  {\mathbf{#1}}
\newcommand{\Lp}[1]            {{\mathsf{L}}_{{#1}}}
\newcommand{\lp}[1]             {{\ell}_{#1}}
\newcommand{\normt}[3]{ {\| {#1} \|}_{ {\Lp{#2}} \rob{#3}}}
\newcommand{\df}[3]			   {{\partial^{(#1)}_{#3}} #2}

\newcommand{\V}[2]             {\mathsf{V}_{#1}^{#2}}
\newcommand{\B}[1]             {\beta_{#1}}
\newcommand{\K}[1]             {\mathcal{K}_{#1}}

\newcommand{\euler}[2]     {\mathcal{E}_{#1,#2}}

\newcommand{\PMs}				{\mathsf{P}}
\newcommand{\QMs}				{\mathsf{Q}}

\newcommand{\Aj}				  {\mathbf{G}^{\sqb{j}}}
\newcommand{\Bj}		       	  {\mathbf{V}^{\sqb{j}}}
\newcommand{\Rj}	      	      {\mathbf{R}^{\sqb{j}}}
\newcommand{\Gj}	      	      {\mathbf{A}^{\sqb{j}}}
\newcommand{\zj}	      	       {\mathbf{z}}

\newcommand{\abs}[1]			{\left| #1\right|}
\newcommand{\inner}[2] 			{\langle {#1,#2} \rangle}

\newcommand{\zm}[2]			   {\xi_{#1}^{#2}}

\newcommand{\mse}[2]		{\EuScript{E}{\rob{\mat{#1},\mat{#2}}}}
\newcommand{\msep}[3]	  {\EuScript{E}{({\mat{#1},\mat{#2};#3})}}
\newcommand\subfig[3]	    {Fig.~\ref{#1} (${\mathsf{#2}}_{#3}$)}
\newcommand{\fe}[1]				{\left[\kern-0.30em\left[#1\right]\kern-0.30em\right]}
\newcommand{\ft}[1]				{\left[\kern-0.15em\left[#1\right]\kern-0.15em\right]}
\newcommand{\flr}[1]			{\left\lfloor #1 \right\rfloor}

\newcommand{\MO}[1]		     {\mathscr{M}_\lambda ({#1} )}
\newcommand{\dfo}[2]          {\Delta^{(#1)} #2}

\renewcommand\bar\underline

\newcommand{\pMs}			{\mathbf{p}}
\newcommand{\tpMs}		   {\mathbf{\widetilde{p}}}
\newcommand{\qMs}			{\mathbf{q}}
\newcommand{\tqMs}		   {\mathbf{\widetilde{q}}}
\newcommand{\QO}[1]		   {\mathscr{Q}_\lambda \rob{{#1}}}
\newcommand{\G}[1]			  {\mat{G}({#1})}
\newcommand{\Ac}[1]          {\mathcal{A}_{#1}}
\newcommand{\Bc}[1]          {\mathcal{B}_{#1}}

\usepackage[many,breakable]{tcolorbox}

\tcbset{
	left = 0pt,
	top = 0pt,
	bottom = 0pt,
	breakable,
	before skip=0.2cm,
	after ={\parindent0em},
	colback=yellow!10!white, colframe=red!50!black, 
	highlight math style= {enhanced, 
		colframe=red,colback=red!10!white,
		boxsep=0pt,
}}

\newtheorem{theorem}{Theorem}
\tcolorboxenvironment{theorem}{
  colback=blue!5,
  boxrule=1pt,
  boxsep=2pt,
  oversize=2pt,
  sharp corners,
  left=2pt,right=2pt,top=2pt,bottom=2pt,
  before skip=\topsep,
  after skip=\topsep,
  width = 0.8\linewidth,
}

\newtheorem{lemma}{Lemma}
\tcolorboxenvironment{lemma}{
  colback=green!5,
  boxrule=1pt,
  boxsep=1pt,
  left=2pt,right=2pt,top=2pt,bottom=2pt,
  oversize=2pt,
  before skip=\topsep,
  after skip=\topsep,
  width = 0.9\linewidth,
}

\newtcbox{\abox}[1][brown]{on line,
	arc=0pt,
	colback=#1!10!white,
	colframe=#1!50!black,
	arc=0pt,
	outer arc=0pt,
	top=1pt,
	bottom=0.5pt,
	left=0mm,
	right=0mm,
	leftrule=0pt,
	rightrule=0pt,
	toprule=0.3mm,
	bottomrule=0.3mm,
	boxsep=0.1mm
}

\makeatletter
\def\moverlay{\mathpalette\mov@rlay}
\def\mov@rlay#1#2{\leavevmode\vtop{
		\baselineskip\z@skip \lineskiplimit-\maxdimen
		\ialign{\hfil$\m@th#1##$\hfil\cr#2\crcr}}}
\newcommand{\charfusion}[3][\mathord]{
	#1{\ifx#1\mathop\vphantom{#2}\fi
		\mathpalette\mov@rlay{#2\cr#3}
	}
	\ifx#1\mathop\expandafter\displaylimits\fi}
\makeatother

\usepackage{titlesec}
\titlespacing{\subsubsection}{0em}{1em}{0.2em}[1em] 

\titleformat{\subsection}[runin]{\normalfont\bfseries}{}{0em}{\thesubsection:~}

\titleformat{\subsubsection}[runin]{\normalfont\it}{}{0em}{\thesubsubsection:~}

\usepackage{scalefnt}
\newcommand{\pbox}[2]     		  {{\colorbox{black}{\color{white}{\sf \scalefont{0.9}{#1#2}}}}}

\begin{document}
\onecolumn

\title{\scalefont{0.8}
Unlocking Off-the-Grid Sparse Recovery \\ with Unlimited Sensing:\\
Simultaneous Super-Resolution in Time and Amplitude
}

\author{Ruiming Guo and Ayush Bhandari

\thanks{
This work is supported by the European Research Council's Starting Grant for ``CoSI-Fold'' (101166158) and UK Research and Innovation council's FLF Program ``Sensing Beyond Barriers via Non-Linearities'' (MRC Fellowship award no.~MR/Y003926/1). Further details on {Unlimited Sensing} and materials on \textit{reproducible research} are available via  \href{https://bit.ly/USF-Link}{\texttt{https://bit.ly/USF-Link}}.}

\thanks{The authors are with the Computational Sensing and Imaging Lab (CSIL) at the Dept. of Electrical and Electronic Engg., Imperial College London, South Kensington, London SW7 2AZ, UK. (Email: \texttt{\{ruiming.guo,a.bhandari\}@imperial.ac.uk} or \texttt{ayush@alum.mit.edu}).}
}

\markboth{IEEE Journal of Selected Topics in Signal Processing,~Vol.~XX, No.~X, Oct~2025}
{Guo \MakeLowercase{\textit{et al.}}: IEEE Journal of Selected Topics in Signal Processing}

\maketitle

\vspace{1cm}

{
\centering

\color{blue} IEEE Journal of Selected Topics in Signal Processing (to appear).

}

\vspace{1cm}

\begin{abstract}
The recovery of Dirac impulses, or spikes, from filtered measurements is a classical problem in signal processing. As the spikes lie in the continuous domain while measurements are discrete, this task is known as super-resolution or off-the-grid sparse recovery. Despite significant theoretical and algorithmic advances over the past decade, these developments often overlook critical challenges at the analog–digital interface. In particular, when spikes exhibit strong-weak amplitude disparity, conventional digital acquisition may result in clipping of strong components or loss of weak ones beneath the quantization noise floor. This motivates a broader perspective: super-resolution must simultaneously resolve both amplitude and temporal structure. Under a fixed bit budget, such information loss is unavoidable. In contrast, the emerging theory and practice of the Unlimited Sensing Framework (USF) demonstrate that these fundamental limitations can be overcome. Building on this foundation, we demonstrate that modulo encoding within USF enables digital super-resolution by enhancing measurement precision, thereby unlocking temporal super-resolution beyond conventional limits. We develop new theoretical results that extend to non-bandlimited kernels commonly encountered in practice and introduce a robust algorithm for off-the-grid sparse recovery. To demonstrate practical impact, we instantiate our framework in the context of time-of-flight imaging. Both numerical simulations and hardware experiments validate the effectiveness of our approach under low-bit quantization, enabling super-resolution in amplitude and time.

\end{abstract}

\begin{IEEEkeywords}
Computational sensing, off-the-grid, super-resolution, sparse recovery, imaging and unlimited sensing.
\end{IEEEkeywords}

\newpage
\tableofcontents

\newpage
\listoffigures

\listoftables

\newpage

\setstretch{1.1}

\section{Introduction}
\label{sec:intro}

Consider the problem of recovering echoes of backscattered light, where each pulse-echo corresponds to an object at a specific depth. Since light travels roughly 30 centimeters (cms) in one nanosecond (ns), resolving objects separated by just a cm requires an analog-to-digital converter (ADC) capable of sub-ns temporal resolution; a requirement that is technologically prohibitive. Although this particular example relates to light-based 3D imaging \cite{Bhandari:2022:Book}—a scenario discussed later in this paper—similar challenges also arise in widespread areas such as radar systems \cite{DeFigueiredo:1982:J}, DNA profiling \cite{Li:2000:J}, optical coherence tomography \cite{Seelamantula:2014:J}, terahertz sensing \cite{RedoSanchez:2016:J}, super-resolution microscopy \cite{Denoyelle:2019:J}, and radio-astronomy \cite{Pan:2017:J}.

\begin{SCfigure}[][h]
\centering
\includegraphics[width=0.72\linewidth]{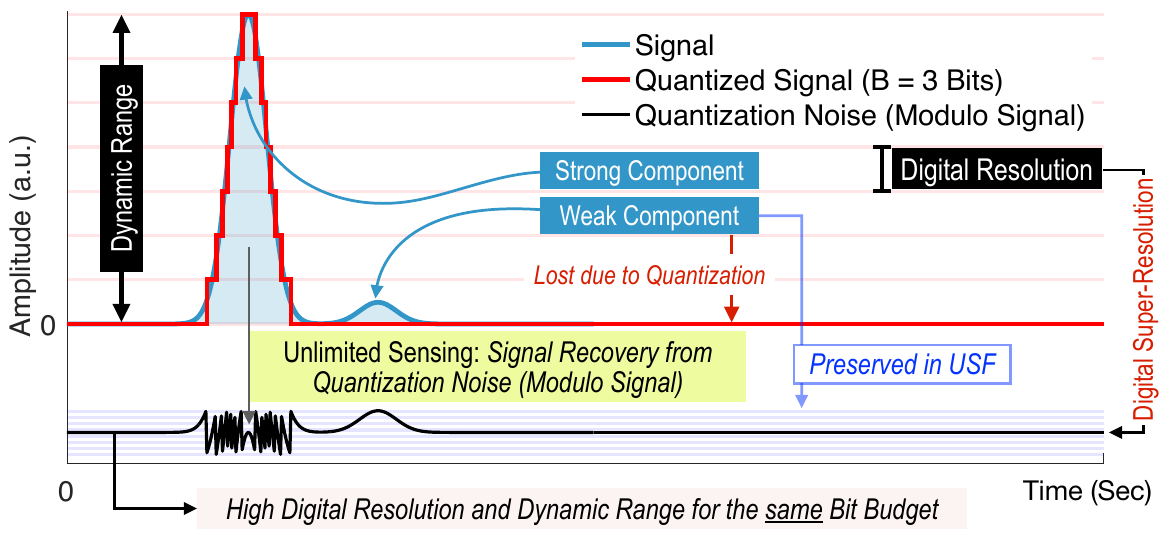}
\caption{
Conventional systems with fixed digital resolution cannot achieve simultaneous amplitude and temporal super-resolution because weak signals are obscured by coarse quantization and rendered unresolvable. The \usf addresses this limitation by enabling digital super-resolution, which enhances measurement precision and facilitates super-resolution in both amplitude and time (see \fig{fig:curve}).
}
\label{fig:WS}
\end{SCfigure}

In scientific and engineering disciplines, this challenge is frequently termed the super-resolution (\SR) problem, aptly named because it involves recovering \emph{fine-scale} features from \emph{coarse-scale} measurements. Mathematically, the problem can be abstracted as recovering sparse spikes (or Dirac masses),
\begin{equation}
\label{eq:SRF}
s_K\rob{t} = \sum\limits_{k = 0}^{K - 1} {\gammar \sqb{k} \delta  \rob{{t - \taur\sqb{k}} } }, \quad 
\{\gammar, \tau\} \in \mathbb{R}
\end{equation}
from filtered measurements,
\begin{align}
g[n] & = g(nT), \qquad n=0, \ \ldots,N-1, \quad T>0 \notag\\
\gr\left( t \right) &\EQc{eq:SRF} \rob{s_K*\psi}\rob{t}  \equiv \sum\limits_{k = 0}^{K - 1} {\gammar \sqb{k}\fr  \rob{{t - \taur\sqb{k}} } }
\label{eq:g}
\end{align}
where $\fr$ represents a smooth kernel and $T$ is ADC's temporal resolution. The goal of \SR is to recover the \emph{off-the-grid}, continuous-time signal $s_K(t)$ from the discrete samples $g[n]$, which lie on a \emph{coarse temporal grid} determined by $T$.

One of the earliest solutions to this problem dates back to the seminal work of de Figueiredo \& Hu (1982) \cite{DeFigueiredo:1982:J}, who showed that $2K$ measurements suffice to recover a continuous-time, $K$-sparse signal, $s_K(t)$ using Prony’s method—a foundational approach in spectral estimation. This method has since underpinned key developments in areas such as \emph{spike deconvolution} \cite{Li:2000:J,Batenkov:2020:J}, \emph{finite-rate-of-innovation} (FRI) sampling \cite{Vetterli:2002:J} and \emph{sub-Nyquist sampling} \cite{Gedalyahu:2010:J}. Over the past decade, there has been a renewed interest in \emph{bridging finite-dimensional measurements} with \emph{infinite-dimensional solution space}. Notable contributions in this direction include the work of Bredies et al. \cite{Bredies:2012:J}, Candès \& Fernandez-Granda \cite{Candes:2013:J}, Tang et al. on compressed sensing off the grid \cite{Tang:2013:J}, and the total variation (TV) regularization approach by Duval \& Peyré \cite{Duval:2014:J}, among others. For a comprehensive overview, we refer the reader to the excellent survey by Chi \& Ferreira Da Costa \cite{Chi:2020:J}.

\subsection{What’s Missing in Previous Art?} 
In practice, ADCs bridge analog and digital domains but are \uline{fundamentally constrained} by a trade-off: for a fixed bit-budget, one cannot simultaneously achieve high \emph{dynamic range} (DR) and high \emph{digital resolution} (DRes). Improving one typically compromises the other. Prioritizing DR results in coarse quantization, causing (i) weak signals to be buried beneath the noise floor in the presence of strong components, as illustrated in \fig{fig:WS}, and (ii) degradation in algorithmic \SR performance due to elevated quantization noise. Conversely, optimizing for DRes leads to clipping of high DR or \HDR signals, causing irreversible information loss and breakdown of \SR methods \cite{Weinstein:2013:J}. Although \SR is well explored, its hardware validation is limited, and the information loss at the analog–digital interface remains under-explored due to high DRes assumptions. For current ADCs, capturing \HDR signals at high-DRes (\HDRes) is {\bf fundamentally impossible}; a key limitation for practical \SR. This motivates a wider perspective: \emph{super-resolution must simultaneously resolve both amplitude and temporal structure}--an ability that conventional systems inherently lack.

Increasing bit depth is unsustainable, as ADC power consumption grows \emph{exponentially} with the number of bits, but only \emph{linearly} with the sampling rate \cite{Walden:1999:J}. As a result, \emph{oversampling} emerges as a more power-efficient alternative. This trade-off becomes especially critical in high-dimensional imaging, where the data volume per pixel/voxel is large, making high bit-budgets impractical due to power and storage constraints, \eg data tensors in Time-of-Flight (ToF) imaging \cite{Bhandari:2016:J,Bhandari:2020:J,Guo:2025:J}.

\subsection{The USF Breakthrough.} The Unlimited Sensing Framework (USF) introduced in \cite{Bhandari:2017:Cb,Bhandari:2020:Ja,Bhandari:2021:J} \uline{breaks the fundamental trade-off between dynamic range (DR) and digital resolution (DRes)} in conventional ADCs by demonstrating that, for a fixed bit-budget, one can simultaneously achieve \HDR signal capture with \HDRes. USF is built on the mathematical insight that, for smooth signals \cite{Bhandari:2020:Ja,Bhandari:2020:C,Beckmann:2022:J,Beckmann:2024:J}, \emph{the integer part can be recovered from the fractional part}. In engineering terms, the integer part corresponds to the conventionally quantized signal, while the fractional part represents the quantization noise, which USF treats as an informative signal representation rather than a distortion artifact. 

The strength of USF arises from hardware–algorithm co-design. 
With folding parameter, $\lambda>0$, the hardware employs an analog-domain modulo non-linearity \cite{Bhandari:2021:J}, defined as:
\begin{equation}
\label{eq:map}
\mathscr{M}_{\lambda}: g \mapsto 2\lambda\left(\fe{\frac{g}{2\lambda}+\frac{1}{2}}-\frac{1}{2}\right), \quad \ft{g}=g-\flr{g},
\end{equation}
where $\flr{g}=\sup\{k\in\mathbb{Z}\mid k\leqslant g\}$ denotes the integer part of $g$. Since $|\mathscr{M}_{\lambda}(g)|\leqslant\lambda$, sampling and quantizing $\mathscr{M}_{\lambda} (g)$ avoids clipping and yields ``digital super-resolution" or \HDRes (see \fig{fig:WS}). Combined with advanced recovery algorithms, this enables new capabilities. Hardware validation of USF has verified a 60-fold DR extension in practice \cite{Zhu:2024:Ca}, along with at least 10 dB enhancement in DRes for modulo tomography \cite{Beckmann:2024:J,Beckmann:2022:J} and radars \cite{Feuillen:2023:C}. Such improvements are critical for advanced modulation formats like 1024-QAM \cite{Liu:2023:J}, full-duplex communication \cite{Liu:2025:J} and \SR spectral estimation \cite{Guo:2024:J,Guo:2025:Cc}.

\begin{figure*}[!t]
 \centering
  \captionsetup{width=.85\linewidth}
\includegraphics[width=0.8\linewidth]{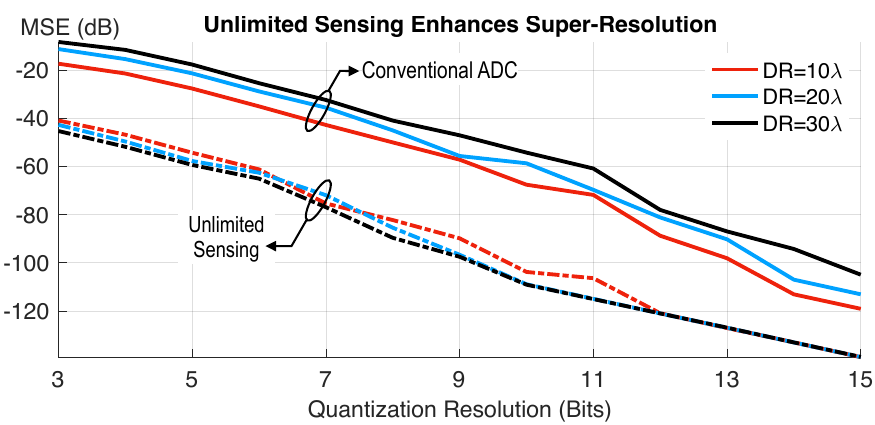}
\caption{Performance evaluation across varying dynamic ranges (DR) and quantization bits. The measurements follow the model $g[n] = \Gamma[1]\psi(nT - \tau[1]) + \Gamma[2]\psi(nT - \tau[2])$ (see \eqref{eq:g}), involving a weak-strong mixture with an amplitude ratio of $\abs{\Gamma \sqb{1}/\Gamma \sqb{2}} = 10$. The detailed experimental setup is described in \secref{subsec:HDR}. We compare the recovery mean-squared error (MSE) between a conventional ADC and \usf across quantization levels ($\sqb{3,15}$) and input DR values ($\norm{g}_{\Lp{\infty}} = \{10, 20, 30\}\lambda$). In all scenarios, \usf consistently achieves a performance gain of at least $30$\,\dB \ and demonstrates robustness to input DR variations.}
\label{fig:curve}
\end{figure*}

\subsection{Challenges and Related Work.}
Achieving \emph{off-the-grid} recovery or super-resolution (\SR) from folded samples gives rise to a new class of inverse problems, which we refer to as \ussr. This topic is still in its nascent stages. 

The first attempt at formulating and addressing \ussr appeared in \cite{Bhandari:2018:Ca}, which provided theoretical recovery guarantees for periodic-bandlimited kernels ($\psi$ in \eqref{eq:g}) under idealized conditions. A subsequent approach in \cite{Bhandari:2022:J} leveraged Fourier-domain recovery using the same class of kernels to directly estimate spike locations $s_K(t)$. This method offered both \emph{sampling guarantees} and a \emph{proof-of-concept hardware validation}, demonstrating approximately $9.5\lambda$ HDR recovery. However, it is susceptible to spectral leakage and cannot handle the broader class of smooth, non-bandlimited kernels that commonly arise in practical applications and form the focus of this work (see \fig{fig:clipping} and \fig{fig:HW_14}).

In \cite{Mulleti:2024:J}, the authors tackle \ussr by relying on prior knowledge of the input to satisfy Itoh’s criterion \cite{Itoh:1982:J}. 
One key advantage of the USF is that it decouples sampling rate from dynamic range (DR) and relaxes Itoh’s restriction, which imposes strict DR constraints (see Fig.~2, \cite{Bhandari:2020:J}). This leads to limited \HDR performance in \cite{Mulleti:2024:J}, as evidenced by hardware results achieving only $2.8\lambda$ recovery (see Fig.~9, \cite{Mulleti:2024:J}). 

Until now, recovery guarantees in all prior works were tied to the sampling rate. In \cite{Guo:2023:Ca}, the authors demonstrated that sparse signals could be recovered \emph{independently of the sampling rate} using a dual-channel architecture; however, similar to \cite{Bhandari:2018:Ca}, this approach assumes an idealized setting. 
A one-bit quantization scheme with time-varying thresholds for sparse recovery \SR was proposed in \cite{Eamaz:2023:C}, based on periodic-bandlimited kernels and idealized folding conditions. While conceptually interesting, this approach relies on highly specialized hardware and ADC architectures, resulting in a niche problem setup that falls outside the scope of \SR and \ussr, as considered in prior works \cite{Bhandari:2018:Ca,Bhandari:2022:J,Mulleti:2024:J,Guo:2023:Ca}.

\subsection{Motivation.} Prior works have not  recognized or leveraged \ussr to enhance temporal resolution via digital super-resolution or increase measurement precision, essential for achieving simultaneous amplitude and temporal super-resolution. This capability is crucial under low-resolution sampling. Two key limitations arise: theoretical models for practical, non-bandlimited kernels ($\psi$ in \eqref{eq:g}) are lacking, and robust algorithms for sparse recovery under coarse quantization are insufficient.

\subsection{Contributions.} This paper addresses the aforementioned limitations and advances the state-of-the-art in off-the-grid super-resolution \cite{Chi:2020:J}. Advocating that \emph{USF unlocks Super-Resolution} in both amplitude and time, the core contribution lies in the algorithmic realization and hardware-level validation of \ussr, even under stringent quantization constraints—as low as 3 bits (see \fig{fig:curve}). To achieve this, we develop a \emph{cohesive framework} that integrates theoretical guarantees, robust algorithm design, and practical hardware experimentation.

\begin{enumerate}[leftmargin = 2em, label = {\ding{111}}, itemsep = 5pt, topsep = 5pt, labelsep = 0.5em]
\item \textbf{Theory.} In view of broader classes of functions, we consider \ussr with non-bandlimited kernels that can not be handled with previous theory. Building on this, we establish theoretical guarantees for sparse signal recovery from modulo-folded samples (see \thmref{thm:1}). 

\item \textbf{Algorithm.} We develop an empirically robust algorithm designed to tolerate {measurement} distortions and hardware non-idealities. Our non-convex optimization {framework} accurately estimates sparse spikes, enabling \HDR signal recovery ($30\lambda$ in \fig{fig:curve}) while mitigating challenges such as spectral leakage, offering clear benefits over previous art \cite{Bhandari:2022:J}, see \fig{fig:clipping}.

\item \textbf{Experimental Validation and Benchmarking.} To bridge the gap between theory and practice, we implement our framework in a time-of-flight (ToF) imaging setup \cite{Bhandari:2016:J,Bhandari:2020:J,Guo:2025:J}. The method is benchmarked under varying conditions using hardware-acquired ToF data, digitized via modulo ADCs \cite{Bhandari:2021:J}. These experiments—with centimeter-level inter-object separations—demonstrate the robustness and \SR capability of our approach, achieving up to $23\lambda$ \HDR recovery (see \fig{fig:HW_14}).
\end{enumerate}

\bpara{Notation.} The set of integer, real, and complex-valued numbers are denoted by $ \mathbb{Z}, \mathbb{R}$ and $\mathbb{C}$, respectively. 
The set of $N$ contiguous integers is denoted by $\id{N} = \{0,\cdots, N-1\}, N\in \mathbb{Z}^{+}$. 
Continuous functions are written as $g\rob{t}, t\in \mathbb{R}$; their discrete counterparts are represented by 
$g\left[ n \right] = {\left. {g\left( t \right)} \right|_{t = nT}}$,
$n\in \mathbb{Z}$ where $T  > 0$ takes the role of sampling period. 
Vectors and matrices are written in bold lowercase and uppercase fonts, such as $\mathbf{g} = [g[0],\cdots,g[N-1]]^{\transp} \iR^{N}$ and $\mat{G} = [g_{n,m}]_{n\in\id{N}}^{m\in\id{M}} \iR^{N\times M}$.
The $\Lp{p}\rob{\R}$ space equipped with the $p$-norm or $\normt{\cdot}{p}{\R}$ is the standard Lebesgue space. For instance, $\Lp{1}$ and $\Lp{2}$ denote the space of absolute and square-integrable functions, respectively. Spaces associated with sequences are denoted by $\ell_{p}$. 
The max-norm $(\Lp{\infty})$ of a function is defined as, $\norm{g}_{\Lp{\infty}} = \inf \{c_0 \geqslant 0: \left|g\rob{t}\right| \leqslant c_0 \}$; for sequences, we use, $\norm{g}_{\lp{\infty}} = \max_{n} \left| g \sqb{n} \right|$.
The $\Lp{2}$-norm of a function is defined as, $\norm{g}_{\Lp{2}} = \sqrt{\int \left|g\rob{t}\right|^{2} d t}$ while for sequences, we have, $\norm{g}_{\lp{2}} = \sqrt{\sum_{n =0}^{N-1} \left| g \sqb{n} \right|^{2}}$.
The inner-product of two functions $f,g\in \Lp{2}$ is defined as, $\inner{f}{g} = \int f(t)\overline{g (t)} dt$ while for sequences, we have, $\inner{\mat{f}}{\mat{g}} = \sum_{n =0}^{N-1} f[n] \overline{g [n]}$. 
The vector space of polynomials with complex coefficients and degrees less than or equal to $K$ is denoted by $P_{K}$, for instance, $\QMs (z) = \sum\nolimits_{k=0}^{K} h_{k} z^{k} \in P_{K}$. 
The $\L$-order derivative of a function is denoted by $\df{\L}{g}{t} \left( t \right)$. The space of $\L$-times differentiable, real-valued functions is denoted by $C^{\L}\rob{\R}$. For sequences,  the first-order finite difference is denoted by $(\Delta g)\sqb{n} = g[n+1]- g\sqb{n}$.
For any function $g\in \Lp{1}$, its Fourier Transform is defined by $\widehat{g} (\omega)  = \int g\rob{t} e^{-\jmath \omega t} dt$. For sequences, the Discrete Fourier Transform (DFT) of a sequence $\mat{g}\in \lp{1}$ is denoted by $\widehat{g}[m] = \sum\nolimits_{n=0}^{N-1} g\sqb{n} e^{-\jmath \frac{2\pi n }{N}m}$. 
Let $\mathbf{W}_{N}^{M}$ denote the $N\times M$ Vandermonde matrix $\mathbf{W}_{N}^{M}=\bigl[ \zm{N}{n\cdot m}  \bigr]_{n\in\id{N}}^{m\in\id{M}}, \ \zm{N}{n} = e^{\jmath\frac{2\pi n}{N}}$. 
The short hand notation for diagonal matrices is given by $\dk \rob{\mat{h}}$ with $\sqb{\dk \rob{\mat{h}}}_{k,k} = \sqb{\mat{h}}_{k\in\id{K}}$. 
We use $\rob{f\circ g}\rob{t}= f\rob{g\rob{t}}$ to denote function composition. 
The mean-squared error (MSE) between $\mat{x}, \mat{y} \iR^N$ is given by $\mse{x}{y} = \frac{1}{N}\sum\nolimits_{n=0}^{N-1} \left|x\sqb{n} -y\sqb{n} \right|^{2}$.

\newpage
\section{Exact Sparse Spike Recovery}
\label{sec:method}

\subsection{Problem Formulation.} 
Let $\gr\rob{t} \gg \lambda$ be a \HDR sparse signal in \eqref{eq:g}. 
Let $\yr\rob{t} = \MO{\gr\rob{t}}$ be the folded version of $\gr\rob{t}$ as in \eqref{eq:map}, then uniform sampling of the LDR, continuous-time signal $\yr\rob{t}$ results in,
\begin{equation}
\label{eq:filtered samples}
\yr \sqb{n} = {\left. {\yr\left( t \right)} \right|_{t = nT}}=  \MO{ \gr\rob{nT} }, \ \ n\in\id{N}
\end{equation}
where $T>0$ is the sampling step. 
Given $\{\yr\sqb{n}\}_{n\in\id{N}}$ and $\{\fr\sqb{n}\}_{n\in\id{N}}$, our goal is to recover $\sr \rob{t}$.
In this section, we conduct the theoretical analysis in noiseless case while taking \emph{noise and quantization} into consideration in \secref{sec:alg}.

\subsection{Tools from Approximation Theory.} 
Since $\MO{\cdot}$ transforms a smooth function into a piecewise smooth one \cite{Bhandari:2020:Ja}, our recovery approach hinges on non-linear filtering of smooth and non-smooth components. Consequently, analyzing the smoothness of the underlying functions becomes central to our method. In this regard, norm-interpolation inequalities play a key role. In particular, our development relies on the Kolmogorov–Landau inequality \cite{Kolmogorov:1949:J}, a focal point of mathematical analysis from the 1950s--80s. The central idea is that, given certain $\Lp{p}$–norms of an operator (\eg\ derivatives), intermediate norms can be estimated via interpolation. The following theorem provides a quantitative formulation.

{
\centering 

\begin{theorem}[Kolmogorov \cite{Kolmogorov:1949:J}]
\label{thm:KLI}
Let $\fr$ be an $L$--times differentiable function defined on the set $\mathbb{T} \subseteq \mathbb{R}$, with bounded derivatives. Then, for $l = 1, \ldots, L-1$, the function $\fr$ satisfies
\begin{equation}
\label{eq:NIE}
\| \df{l}{\fr}{t} \|_{\Lp{\infty}(\mathbb{T})} \leq \mathscr{C}_{l,L} \left( \| \fr \|_{\Lp{\infty}(\mathbb{T})}^{\,1-l/L} \right) \left( \| \df{\L}{\fr}{t} \|_{\Lp{\infty}(\mathbb{T})}^{\,l/L} \right). 
\end{equation}
\end{theorem}

}

Kolmogorov’s pioneering work established the sharp constants $\{\mathscr{C}_{l,L}\}_{l=1}^{L-1}$ for $\mathbb{T} = \R$ through construction of extremal functions that attain \eqref{eq:NIE}. In particular, 
\begin{equation}
\label{eq:clL}
\mathscr{C}_{l,L}= \frac{\K{\L-l}}{\K{\L}^{1-l/\L}}, \quad
\K{\L} = \norm{\euler{m}{\L}}_{\Lp{\infty}}
\end{equation}
where $\euler{m}{\L}, L\geqslant 0$, is the \emph{perfect Euler spline} of order-$\L$,
\begin{align}
\euler{m}{\L}\rob{t} & = \frac{4}{\pi m^{\L}} \sum\limits_{p\geqslant0} \frac{\sin( (2p+1)m t - \pi \L/2)}{(2p+1)^{\L+1}}, \quad m\geqslant 1 \notag\\ 
\mbox{and} \ \K{\L} 
& = \frac{4}{\pi} \sum_{p=0}^{\infty}  \rob{\frac{(-1)^p}{2p + 1}}^{L+1}
\mbox{{(Bohr-Favard constant)}}.
\label{eq:KL}
\end{align}

\subsection{Main Result: Sampling Theorem and  Recovery Algorithm.} 
We propose a time-domain approach that uses digital \SR along with amplitudes to achieve temporal \SR. This enhances temporal \SR capabilities while reducing sampling costs, a critical advantage lacking in conventional \SR methods that assume infinite-resolution measurements. The smoothness of the kernel $\fr$ enables amplitude shrinkage in the high-order difference domain, allowing for the separation of the integer part $\gr - \MO{\gr}$ via non-linear filtering. Spectral fitting then retrieves $\sr\rob{t}$ perfectly.

\bpara{Mathematical Model and Properties of the Kernel.} In developing an off-the-grid \usf method, an effective starting point entails abstracting properties of $\fr$ shared across various data modalities. Our observation, based on various experimentally calibrated kernels used in practice, suggests that these kernels exhibit common mathematical properties \cite{Guo:2025:J}: 
\begin{enumerate}
\item \textit{Time-concentration}: $\fr (t) \approxeq 0,\; t \not\in \interval \subset \mathbb{R}$ where $\interval$ is finite, contiguous time interval.
\item \textit{Smoothness}: $\fr \in C^{\L}\rob{\R}$ is $\L$-times differentiable.
\end{enumerate}

Consequently, we model the kernel as a function that belongs to a shift-invariant space (SIS) \cite{Deboor:1994:J} or $\fr \in \V{\gamma}{}$, where, $\V{\gamma}{\gf} = 
\mathrm{span}
\left\{ \gf \rob{  {t}/{\gamma} - l } \right\}_{l \in \Z}$ 
and where $\gf$ is the generator of the SIS. Any $\fr \in \V{\gamma}{\gf}$ can be represented as, 
\begin{equation}
\label{eq:basis}
\fr \rob{t} = \sum\limits_{l\in\Z}  \coef\sqb{l} \gf \rob{\frac{t}{\gamma} - l } , \quad \gamma>0
\end{equation}
where $\gamma$ regulates the density of the grid. This representation generalizes the Shannon-Nyquist representation for bandlimited spaces or $\V{\gamma}{\mathrm{sinc}}$, extending to compactly supported generators and non-bandlimited function spaces.
The representation in \eqref{eq:basis} is stable and unique, whenever the translates $\{\gf(t-k)\}_{k\in\mathbb{Z}}$ constitute an $\Lp{p}$-stable Riesz basis\footnote{The Riesz basis property ensures that the $\Lp{p}$-norm of a function is equivalent to the $\ell_p$-norm of its expansion coefficients in the SIS. In our case, when working with square-integrable functions, the existence of Riesz bounds guarantees that the basis functions belong to $\Lp{2}$ and are linearly independent in the $\ell_2$ sense. Whenever the basis is orthornormal, we have $\Ac{\gf} = \Bc{\gf} =1$ and the Parseval’s relation holds.} \cite{Aldroubi:1994:J}. That is to say, 
$\forall  p\in [1, \infty]$, there exist lower and upper bounds, $0<\Ac{\gf} \leqslant \Bc{\gf} < \infty$ such that, 
\begin{align}
\label{eq:RB}
\ \ \  \Ac{\gf} & =  \inf_{\|b\|_{\ell_p} = 1} \left\| \sum\limits_{l \in \mathbb{Z}} b[l] \, \gf(t-l) \right\|_{\Lp{p}} > 0 
\quad \mbox{and} \quad \\  \Bc{\gf} & =  \sup_{\|b\|_{\ell_p} = 1}  \left\| \sum\limits_{l \in \mathbb{Z}} b[l] \, \gf(t-l) \right\|_{\Lp{p}} < \infty. 
\end{align}

In our work, we choose B-splines \cite{Aldroubi:1994:J} as the SIS generator,
\begin{equation}
\label{eq:spline}
\B{\L}(t) = \frac{1}{L!} \sum_{k=0}^{L+1} \binom{L+1}{k} (-1)^k 
\left( t - k + \frac{L+1}{2} \right)_{+}^{\,L}
\end{equation}
where $L$ is the spline order and $\rob{t}_+^L$ denotes the one-sided power function.  Our choice is motivated by the fact that splines satisfy all of the above properties--- \emph{smoothness}, \emph{time-concentration} (compact support) and \emph{stable Riesz basis} \cite{Unser:2000:J}.

Let $\V{\gamma}{L} = \mathrm{span}\left\{ \B{\L}\rob{  {t}/{\gamma} - l } \right\}_{l \in \Z}$.
In the next result, we show that $\fr \in \V{\gamma}{L}$ allows for analytical approximation of the kernel $\fr$, setting the foundation for our recovery approach. 

{
\centering

\begin{lemma}
\label{lemma:1}
Let $\fr\in \V{\gamma}{\L} \cap \Lp{\infty}\rob{\R}$ with $\fr (t) = 0,\; t \not\in \interval \subset [0,\tau)$ with $\K{L}$ in \eqref{eq:KL}.
Then, $\fr$ satisfies the approximation property, 
\begin{equation}
\label{eq:BL}
\mathop {\min}_{\{\ai\}} \frac{1}{\tau}\int\limits_{0}^{\tau} \abs{ \fr \rob{t} -   \sum\limits_{\abs{i} \leqslant I}  \ai e^{\jmath {2i \pi t}/{\tau}} }^{2} dt  \leqslant \rho_{L} \quad
\mbox{where} \quad
\rho_{L} \DE \rob{ \frac{  \tau   }{2\gamma  } }^{2\L}  \frac{2\K{\L}^{-2} \norm{\fr}^{2}_{\Lp{\infty}}}{(2\L - 1) I^{2\L - 1}}.
\end{equation}
\end{lemma}
}

\begin{proof}
From the compact support property, we have,
\begin{equation}
\notag
\fr \rob{t}= \sum\limits_{i\iZ} \ak e^{\jmath \tfrac{2i \pi t}{\tau}}, \quad \ak = \tfrac{1}{\tau} 
{\langle{\fr \rob{t},e^{\jmath \tfrac{2i \pi t}{\tau}}}\rangle}_{\left[{{0}{,}{\tau}}\right]}.
\end{equation}
Integration by parts gives $ \ak 
=  \kappa_i^{-1} \int_0^\tau \fr(t)\, d (e^{-\jmath \frac{\kappa_i t}{\tau}})$, 
where $\kappa_i = 2\pi i$. Further simplification yields $ \fr(t) e^{-\jmath \frac{\kappa_i t}{\tau}} |_{0}^{\tau} 
- \kappa_i^{-1} \int_0^\tau e^{-\jmath \frac{\kappa_i t}{\tau}}\, d(\fr(t)) $. Since $\fr\in \V{\gamma}{\L}$ and 
$\fr (t) = 0,\; t \not\in \interval \subset [0,\tau)$, we deduce $\fr(t) e^{-\jmath \frac{\kappa_i t}{\tau}}|_{0}^{\tau} =0$ resulting in, 
\begin{equation}
\label{eq:induction}
\ak = \kappa_i^{-1}\, \int_0^\tau \partial_t^{(1)} \fr(t)\,
e^{-\jmath \frac{\kappa_i t}{\tau}} \, dt, \quad \kappa_i = 2\pi i .
\end{equation}
By repeatedly performing integration by parts, we obtain,
\begin{equation}
\label{eq:spline1}
\abs{\ak} = \frac{ \abs{ \int_{0}^{\tau} \df{\L}{\fr}{t}\rob{t}  e^{-\jmath \tfrac{2i \pi t}{\tau}} dt   }}{\rob{2\abs{i} \pi}^{\L}\tau^{-(\L - 1)}} 
\leqslant   
\rob{\frac{  \tau}{{2\abs{i} \pi}}  }^L
\norm{\df{\L}{\fr}{t}}_{\Lp{\infty}}.
\end{equation}
From the above relation, we establish a connection between the Fourier frequency index $i$ and the smoothness parameter $L$ which serves as a proxy for the effective bandwidth. Specifically, $|{\ak}|\propto C^L_{\fr} i^{-L}$ where $C^L_{\fr}= ||{\df{\L}{\fr}{t}}||_{\Lp{\infty}}$.

For entire functions of exponential type, including
$\fr \in \V{\gamma}{\mathrm{sinc}}$, $||{\df{\L}{\fr}{t}}||_{\Lp{p}}$ can be upper bounded in ${\Lp{p}}$-sense by using the \emph{Bern\v{s}te\u{i}n's inequality} \cite{Nikolskii:1975:B} or $||{\df{\L}{\fr}{t}}||_{\Lp{p}} \leqslant \rob{\pi/\gamma}^L ||{{\fr}}||_{\Lp{p}}$. However, in our case, 
$\fr\in \V{\gamma}{\L} \cap \Lp{\infty}\rob{\R}$;
we consider non-bandlimited, spline-spaces. Nonetheless, as shown in \cite{Bhandari:2020:C}\footnote{
The results in \cite{Bhandari:2020:C} are derived from the extremal properties of polynomials, as in: \\[1pt]
\noindent\includegraphics[width=1\textwidth]{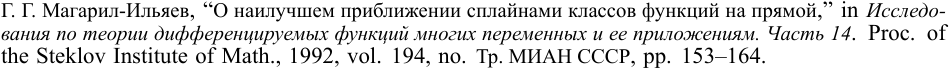}},
using a constant-factor correction to the {Bern\v{s}te\u{i}n's inequality}, it is possible to upper bound $||{\df{\L}{\fr}{t}}||_{\Lp{\infty}}$ using,
\begin{equation}
\label{eq:Maga}
\norm{\df{\L}{\fr}{t}}_{\Lp{\infty}} \leqslant 
\rob{\frac{\pi}{\gamma}}^{\L} 
{\frac{\norm{\fr}_{\Lp{\infty}}}{\K{\L}}}
\xRightarrow{\eqref{eq:spline1}} \; \;
\abs{\ak} \leqslant   \rob{ \frac{  \tau   }{2\gamma \abs{i} } }^{\L} 
{\frac{\norm{\fr}_{\Lp{\infty}}}{\K{\L}}}. 
\end{equation}
Let $\Poly{\ma}{t} =  \sum\nolimits_{\abs{i} \leqslant I}  \ai e^{\jmath {2i \pi t}/{\tau}}$ be a trigonometric polynomial parametrized by $\ma\in\C^{(2I+1)}$. Choosing
$\{\ai = \ak\}_{\abs{i}\leqslant I}$ yields,
\[
{ \fr \rob{t} -  \Poly{\{\ak\}_{\abs{i}\leqslant I}}{t} } = \sum\limits_{\abs{i}>I} \ak e^{\jmath {2i \pi t}/{\tau}}. 
\]
Let $\msep{\fr }{\RMs}{\ma} = \frac{1}{\tau}
|| \fr \rob{t} -  \Poly{\ma}{t}\fr \rob{t} ||^{2}_{\Lp{2}\rob{\R\rob{
\sqb{0,\tau}}}}$ denote the mean-squared error, then by using \eqref{eq:Maga}, we obtain,
$\msep{\fr }{\RMs}{\{\ak\}_{\abs{i}\leqslant I}} 
\leqslant  ( \frac{  \tau   }{2\gamma  } )^{2\L}  \frac{2 \K{\L}^{-2} \norm{\fr}^{2}_{\Lp{\infty}}}{(2\L - 1) I^{2\L - 1}}$, as derived below,
\begin{align}
\msep{\fr }{\RMs}{\{\ak\}_{\abs{i}\leqslant I}} 
&= \sum\limits_{\abs{i}>I}  \abs{   \ak }^{2} \notag \\
& \leqslant 
\rob{ \frac{  \tau   }{2\gamma  } }^{2\L}  \sum\limits_{\abs{i}>I}   \frac{{\norm{\fr}^{2}_{\Lp{\infty}}}}{ {i}^{2\L} } \notag \\
&\leqslant  2\rob{ \frac{  \tau   }{2\gamma  } }^{2\L}  \frac{\norm{\fr}^{2}_{\Lp{\infty}}}{\K{\L}^{2}} \int_{I}^{\infty} \rob{\frac{1}{t^{2\L}}}  dt 
= \rob{ \frac{  \tau   }{2\gamma  } }^{2\L}  \frac{2 \K{\L}^{-2} \norm{\fr}^{2}_{\Lp{\infty}}}{(2\L - 1) I^{2\L - 1}}.
\end{align}
By definition, $\mathop {\min}_{\ma} \msep{\fr }{\RMs}{\ma} \leqslant \msep{\fr }{\RMs}{\{\ak\}_{\abs{i}\leqslant I}}$, 
we then derive the desired result in \eqref{eq:BL}.
\end{proof}
From \lemmaref{lemma:1}, we deduce that for large bandwidths, \ie, $\Ofr = \Ifr \tfrac{2\pi}{\tau}$ with $\Ifr \in \mathbb{Z}^{+}$, where $\Ifr \tfrac{2\pi}{\tau}$ is the largest approximated frequency in \eqref{eq:BL}, $\fr$ can be approximated via \eqref{eq:BL}.

\bpara{Recovery Guarantees.} We formalize the theoretical gurantee for spike recovery in the following theorem.

{
\centering 

\begin{theorem}
\label{thm:1}
Let $\gr\left( t \right) = \sum\nolimits_{k = 0}^{K - 1} {\gammar \sqb{k}\fr  \rob{{t - \taur\sqb{k}} } }$ 
where $\fr\in \V{\gamma}{\L} \cap \Lp{\infty}\rob{\R}$ is a known kernel with compact support $\interval$ and assume $(\max_{k} \abs{\taur\sqb{k}}) + \abs{\interval} <\tau$.
Given $\yr \sqb{n} =  \MO{ \gr\rob{nT} }, n\in\id{N}, T= \tau/N$, the folded samples of $\gr$, then the sparse signal $s_K$ in \eqref{eq:SRF}
can be exactly recovered provided, 
\[
T\leqslant \min\rob{ \frac{\tau}{\L + 2\zn K},\; \frac{\gamma}{\pi} \sqrt[h]{\frac{ {\K{\L} } \lambda }{ \K{\L-h}  \norm{ s_K  }_{\mathsf{TV}} \norm{\fr}_{\Lp{\infty}} }}  }
\] where $$h\in\{1,\cdots, L\} \mbox{ and } \zn \geqslant 1.$$
\end{theorem}

}

\begin{proof}

Let $\dfo{h}{\gr}\sqb{n} = \dfo{h-1}{\gr}\sqb{n+1} - \dfo{h-1}{\gr}\sqb{n}$ define the $h$-th order finite difference. Similarly, define the shift-difference operator as $\SD\rob{\cdot} = (\cdot)\rob{t+T} - (\cdot)\rob{t}$. Then, 
\begin{equation}
\label{eq:SD}
\dfo{h}{\gr}\sqb{n} = {\left. { \SD^{(h)} \rob{\gr} \rob{t} } \right|_{t = nT}}, \ \ \SD^{(h)} = \SD \circ \SD^{(h-1)}
\end{equation}
where $\SD^{(1)} = \SD$. 
By definition, $\dfo{1}{\gr}\sqb{n}$ is upper bounded,
\begin{equation*}
\norm{\dfo{1}{\gr}}_{\lp{\infty}} \leqslant \norm{ \SD\rob{\gr} }_{\Lp{\infty}}  \leqslant T \norm{ \df{1}{\gr}{t}  }_{\Lp{\infty}}  .
\end{equation*}
Moreover, for $h=2$, we obtain, 
\begin{equation*}
\norm{ \SD^{(2)} \rob{\gr} }_{\Lp{\infty}} \leqslant T \norm{ \df{1}{  \SD\rob{\gr} }{t}    }_{\Lp{\infty}}  \leqslant T^{2} \norm{ \df{2}{\gr}{t} }_{\Lp{\infty}} .
\end{equation*} 
Consequently, by induction, for arbitrary $h\in\sqb{1,\L}$ we have
\begin{equation}
\label{eq:gd}
\norm{\dfo{h}{\gr}}_{\lp{\infty}} \leqslant \norm{ \SD^{(h)} \rob{\gr} }_{\Lp{\infty}}  \leqslant T^{h} \norm{ \df{h}{\gr}{t}  }_{\Lp{\infty}} .
\end{equation}
Next, we bound $ \dfo{h}{\gr} $ from above
by analyzing
$\| \df{h}{\gr}{t}  \|_{\Lp{\infty}}$. From \eqref{eq:g} and Young's convolution inequality, we deduce that,
\begin{equation}
\label{eq:gderivative}
\norm{ \df{h}{\gr}{t}  }_{\Lp{\infty}} \leqslant 
\norm{ s_K  }_{\mathsf{TV}}
\norm{ \df{h}{\fr}{t}  }_{\Lp{\infty}}, \;\; h\in\sqb{1,\L}
\end{equation}
where $\norm{ s_K  }_{\mathsf{TV}} = \int |s_K\rob{t}|dt$ denotes
the TV-norm\footnote{
Denotes the continuous version of $\ell_1$-norm for absolutely continuous functions and in the case of Dirac measures, $s_K\rob{t} = \sum\nolimits_{k = 0}^{K - 1} {\gammar \sqb{k} \delta  \rob{{t - \taur\sqb{k}} } }$, we have
$\norm{ s_K}_{\mathsf{TV}} = \sum_k{|\gammar[k]|}$ (also see \cite{Candes:2013:J}).}. Through \eqref{eq:gderivative}, we have transferred the problem of upper bounding $\norm{ \df{h}{\gr}{t}  }_{\Lp{\infty}}$ to $\norm{ \df{h}{\fr}{t}  }_{\Lp{\infty}} $. 
Since $\fr\in \V{\gamma}{\L} \cap \Lp{\infty}\rob{\R}$, then, from the \emph{Kolmogorov-Landau Inequality} in Theorem \ref{thm:KLI}, we have, 
\begin{equation}
\label{eq:Kolmogorov}
\norm{ \df{h}{\fr}{t}  }_{\Lp{\infty}} \leqslant 
{\tfrac{\K{\L-h}}{\K{\L}^{1-h/\L}}}
\rob{  \norm{\fr}^{1 - h/\L}_{\Lp{\infty}} } \rob{\norm{\df{\L}{\fr}{t}}^{h/\L}_{\Lp{\infty}}}.
\end{equation}
By plugging \eqref{eq:Maga} into \eqref{eq:Kolmogorov}, we upper bound $\df{\L}{\fr}{t}$:
\begin{align}
\norm{\df{h}{\fr}{t}}_{\Lp{\infty}} 
\leqslant 
\frac{\K{\L-h}}{\K{\L} }   
\rob{\frac{\pi}{\gamma}}^{h}
\norm{\fr}_{\Lp{\infty}} 
\xRightarrow{\eqref{eq:gderivative}} \; \; 
\norm{\df{h}{\gr}{t}}_{\Lp{\infty}} \leqslant  \frac{\K{\L-h}}{\K{\L} }  \rob{\frac{\pi}{\gamma}}^{h}  \norm{ s_K  }_{\mathsf{TV}} \norm{\fr}_{\Lp{\infty}}.
\label{eq:gdiff}
\end{align}
With \eqref{eq:gd} and \eqref{eq:gdiff}, $\| \dfo{h}{\gr} \|_{\lp{\infty}}$ is thus upper bounded by,
\begin{equation}
\label{eq:gde}
\norm{\dfo{h}{\gr}}_{\lp{\infty}}  \leqslant  \frac{\K{\L-h}}{\K{\L} }  \rob{\frac{\pi T}{\gamma}}^{h}  \norm{ s_K  }_{\mathsf{TV}} \norm{\fr}_{\Lp{\infty}}.
\end{equation}
By regulating $T$ in \eqref{eq:gde}, we can shrink the RHS arbitrarily. In particular, choosing, 
\begin{equation}
\label{eq:shrinkage}
T \leqslant \rob{\frac{\gamma}{\pi}} \sqrt[h]{\frac{{\K{\L-h}} \lambda }{{\K{\L} } \norm{ s_K  }_{\mathsf{TV}} \norm{\fr}_{\Lp{\infty}} }} \Longrightarrow  \| \dfo{h}{\gr} \|_{\lp{\infty}} \leqslant \lambda.
\end{equation}

\bpara{Signal Recovery via Non-linear Filtering.} The result in \eqref{eq:shrinkage} enables us to link $\dfo{\L}{\gr}$ with $\dfo{\L}{\yr}$; from the modular decomposition property \cite{Bhandari:2020:Ja,Guo:2023:C}, we have
\begin{equation}
\label{eq:moddec}
\gr = \MO{\gr} + \res, 
\ \ 
\res\rob{t} = \sum\limits_{m=0}^{\Mr-1} \cm u \rob{t-\taum}
\end{equation}  	
where $u \rob{\cdot}$ is the unit step function and, the \emph{unknowns} $\cm\in2\lambda\Z$ and $\taum \in T\Z^{+} \cap [0,\tau)$ parametrize the folds induced by $\MO{\cdot}$, and where $\Mr$ is the number of folds. From the modular arithmetic, 
$\MO{\sum_k a_k} =\MO{\sum_k \MO{a_k}}$, $\| \dfo{h}{\gr} \|_{\lp{\infty}} \leqslant \lambda$ deduced in \eqref{eq:shrinkage} and $\dfo{\L}{\res}\in2\lambda\Z$, we have (also see \cite{Bhandari:2020:Ja}):
\begin{align}
\label{eq:NF}
\MO{\dfo{\L}{\yr}} \equiv
\MO{\dfo{\L}\rob{\gr-\res}} = \dfo{\L}{\gr}.
\end{align}

\bpara{Sparse Recovery via Spectral Fitting.} Having  recovered $\dfo{\L}{\gr}$ via \eqref{eq:NF}, we next address spike retrieval in the difference domain. Let $\hfr\sqb{l}$ denote the Discrete Fourier Transform (DFT) of $\fr\sqb{n}$. By choosing $\ai = \{\ak\}_{ \abs{i}\leqslant \Ifr}$, the analytical representation of $\fr$ in \eqref{eq:BL} establishes a linear mapping between $\hfr\sqb{l}$ and $\akl$:
\begin{equation}
\hfr\sqb{l} 
\EQc{eq:BL} \sum\limits_{\abs{i}\leqslant \Ifr} \ak \sum\limits_{n=0}^{N-1}  e^{\jmath \tfrac{2(i - l) \pi n}{N}} = N \akl.   
\end{equation}
Similarly, it follows that for $\hgr$,
\begin{align}
\label{eq:bgr}
\hgr \sqb{l} &\EQc{eq:g} \sum\limits_{n=0}^{N-1} \sum\limits_{k = 0}^{K - 1} {\gammar \sqb{k}\fr  \rob{{nT - \taur\sqb{k}} } }  e^{\tfrac{-\jmath 2ln\pi}{N}} = \hfr\sqb{l} \sum\limits_{k = 0}^{K - 1} \gammar \sqb{k}  e^{\jmath \tfrac{- 2l \pi  \taur\sqb{k}}{\tau}}  , \ \ l\in\id{\Ifr+1}.
\end{align}
Next, we show that the linear mapping still holds in the high-order finite difference domain. Starting with $\dfo{1}{\gr}\sqb{n} = \gr\sqb{n+1} - \gr\sqb{n}$, from \eqref{eq:SD} and \eqref{eq:BL}, we can write,
\begin{equation}
\dfo{1}{\gr}\sqb{n} \approxeq  \sum\limits_{\abs{i}\leqslant \Ifr} \ak(1 - e^{\jmath \tfrac{2i \pi T}{\tau}} ) e^{\jmath \tfrac{2i \pi t}{(N-1)T}} 
\end{equation}
which follows the assumption that $\tau = NT$ and \eqref{eq:shrinkage}. Hence, performing the DFT on $\dfo{1}{\gr}\sqb{n}$ and $\dfo{1}{\fr}\sqb{n}$, the linear mapping in \eqref{eq:bgr} still applies. By induction, we can derive that the same conclusion still holds as long as {$\tau \gg LT$}.
Re-organizing \eqref{eq:bgr} leads to a sum-of-sinusoids (SoS) model,
\begin{equation}
\label{eq:hsr}	
\hsr \sqb{l} = \frac{\hgr \sqb{l}}{\hfr \sqb{l}}	 = \sum\limits_{k = 0}^{K - 1} \gammar \sqb{k}  e^{\jmath \tfrac{- 2l \pi  \taur\sqb{k}}{\tau}}, \ l\in\id{\Ifr+1} \backslash \{0\}
\end{equation}
where the recovery of $\{\gammar \sqb{k},\taur\sqb{k}\}_{k=0}^{K-1}$ can be solved using Prony's method \cite{Prony:1795:J}:
Let $\f\sqb{l}$ be the annihilation filter with z-transform $\hf\rob{z} = \sum_{l=0}^{K} \f\sqb{l} z^{-l} = \prod_{k=0}^{K-1} (1 - \uk z^{-1}) $ where $\uk = e^{-\jmath { 2 \pi  \taur\sqb{k}}/{\tau}}, k\in\id{K} $. The roots of $\hf\rob{z}$ uniquely determine the spike instants $\{\taur\sqb{k}\}_{k\in\id{K}}$. Then, it suffices to show that $\f$ annihilates the SoS sequence $\hsr$:
\begin{equation}
\notag
\rob{ \f * \hsr }\sqb{l} 
= \sum\limits_{l_1=0}^{K} \f\sqb{l_1} \sum\limits_{k = 0}^{K - 1} \gammar \sqb{k}  e^{\jmath \tfrac{- 2(l - l_1) \pi  \taur\sqb{k}}{\tau}} 
= \sum\limits_{k = 0}^{K - 1} \gammar \sqb{k} e^{-\jmath \tfrac{ 2l  \pi  \taur\sqb{k}}{\tau}} \hf\rob{\uk} = 0.
\end{equation}
In vector-matrix form, the above is written as
\begin{equation}
\label{eq:AF}
\underbrace{\begin{bmatrix}
\hsr\sqb{K+1} &\hsr\sqb{K} &\cdots &\hsr\sqb{1} \\
\hsr\sqb{K+2} &\hsr\sqb{K+1} &\cdots &\hsr\sqb{2} \\ 
\vdots	&\vdots &\ddots &\vdots\\
\hsr\sqb{\Ifr} &\hsr\sqb{\Ifr-1} &\cdots &\hsr\sqb{\Ifr-K} \\ 
\end{bmatrix}}_{\G{\hsr}}
\underbrace{\begin{bmatrix}
\f\sqb{0}\\
\f\sqb{1}\\
\vdots\\
\f\sqb{K}
\end{bmatrix}}_{\mf} = \mathbf{0}
\end{equation}
where $\G{\hsr}$ is a Toeplitz matrix constructed by $\{\hsr\sqb{l}\}$.
Hence, the annihilation filter coefficients $\mf$ can be obtained by solving the linear system of equations, provided that the number of samples is no smaller than that of unknowns, \ie, $\Ifr  \geqslant 2K$. 
Given $\tau=NT$, this implies that $\rob{\tau/T} - L \geqslant \Ifr \geqslant 2\zn K, \zn \geqslant 1$ ($\rob{\tau/T} - L$ is the sample size of $\{\dfo{\L}{\gr}\sqb{n} \}$). Combining with \eqref{eq:shrinkage}, we derive the sampling condition on $T$. 
Then, $s_K$ defined via the unknowns $\{\gammar \sqb{k}, \taur\sqb{k}\}_{k\in\id{K}}$  can be computed by evaluating the zeros of $\hf\rob{z}$ and solving least-squares problem.
Note that, to make \eqref{eq:hsr} hold, the time-duration $\tau$ should satisfy that, 
$\gr (t) = 0,\; t \not\in [0,\tau) \Longrightarrow  \max_{k} \abs{\taur\sqb{k}} + \abs{\interval} <\tau$, so as to avoid truncation\footnote{From a practical viewpoint, violation of this condition implies that the waveform of $\gr$ is not \emph{fully} covered in the observation window $[0,\tau)$, which may compromise the accuracy of off-the-grid signal recovery.}.
\end{proof}

\bpara{Remarks.} The implications of \thmref{thm:1} are twofold.
\begin{enumerate}
\item The \USSR (Fourier–Prony \SR) approach \cite{Bhandari:2022:J} relies on Fourier-domain partitioning, and its digital \SR performance is constrained by spectral leakage (e.g., $\lVert g \rVert_{L_\infty} = 9.46\lambda$ \cite{Bhandari:2022:J}). By contrast, the our method is inherently agnostic to spectral leakage and achieves a  considerable DR extension ($\lVert g \rVert_{L_\infty} = 30\lambda$ in \fig{fig:curve} and $\lVert g \rVert_{L_\infty} = 23.51\lambda$ in \fig{fig:HW_14}).
\item The parameters $N$ and $T$ govern sample redundancy, indicating that oversampling directly improves the accuracy of the measurement model.
\end{enumerate}

\section{Robust Off-the-Grid \ussr}
\label{sec:alg}

We have shown a theoretically guaranteed spike recovery approach in \thmref{thm:1}, offering concurrent \SR in both temporal and range dimensions. Nonetheless, the fragility of the key assumption of ideal folding, \ie, $\cm\in2\lambda\Z$, together with challenges in real-world scenarios may compromise the \HDR and \SR capability of this method. Specifically, challenges we have identified from hardware experiments include, 
\begin{enumerate}[leftmargin=40pt, label = \pbox{C}{\arabic*}] 
\item {\bf Measurement distortion.} Inherent in electronic circuits, data distortion and noise arise from (i) analog-domain folding and (ii) sampling process, leading to a mixture of non-ideal folding $\cm\notin2\lambda\Z$, quantization noise (bounded, uniformly distributed) and thermal noise (unbounded, Gaussian distributed) \cite{Guo:2024:J,Bhandari:2021:J}. The measurement distortion can destabilize the non-linear filtering operation \eqref{eq:NF}.
\item {\bf Numerical Stability.} As the high-order finite difference operation promotes high-pass frequency components, the signal-to-noise ratio degrades as $h$ increases. This imposes the resolution limits of inter-spike separation, particularly under the low-resolution sampling setup.
\end{enumerate}
This necessitates the design of a robust and flexible \SR method which is resilient to measurement noise and folding non-idealities.
To this end, we introduce \sritersis (\viz \SR iterative signal sieving method) in the following section.

\subsection{Towards Robust \SR Spike Estimation.} Here, we capitalize on the \emph{dual-sparsity features} of the folded measurements. Plugging \eqref{eq:g} and \eqref{eq:moddec} into \eqref{eq:SD}, we can write,
\begin{align}
\label{eq:HOM}
\dfo{h}{\yr} \sqb{n} &= \dfo{h}{ \sum\limits_{k = 0}^{K - 1} {\gammar \sqb{k}\fr  \rob{{nT - \taur\sqb{k}} } }  } 
-\dfo{h}{  \sum\limits_{m=0}^{\Mr-1} \cm u \rob{nT-\taum}  } 
\end{align}
where specifically, we concentrate on $h=1$ since in practice, $h>1$ would inherently (i) amplify the noise level, (ii) increase the folding count and (iii) reduce the signal power, degrading the resolution of spike estimation. 

To simplify the notations, we use $\bgr\sqb{n} = \gr\sqb{n+1} - \gr\sqb{n}$ to denote the finite difference. Hence, \eqref{eq:HOM} simplifies to,
\begin{equation}
\label{eq:model}
\byr \sqb{n} = { \sum\limits_{k = 0}^{K - 1} {\gammar \sqb{k}\bfr  \rob{{n - \frac{\taur\sqb{k}}{T}} } }  }  
-  \sum\limits_{m=0}^{\Mr-1} \cm \delta\sqb{n - \nm} 
\end{equation}
where $\nm=\taum/T \in\id{N}$. 
\eqref{eq:model} follows an \emph{additive \SR} formulation where the spike recovery entails separation of two \uline{sparse, additive terms}: 
\begin{enumerate}
\item bandlimited projection of spikes $\bgr$,
\item sparse spike sequence $\bres$.
\end{enumerate}
The model in \eqref{eq:model} shows the ``low-pass + spikes'' characteristics of the measurements $\byr$, allowing for separating different components apart via time-domain sieving strategy. Hence, in view of \eqref{eq:model}, the off-the-grid \usf problem in real-world scenarios can be posed as:
\begin{align}
\label{eq:global problem}
&\mathop {\min}
_{\substack{ 
\{ 
\mgammar, \mtaur,
\mcm,\mnm
\}
}}
\norm{  \mbyr - \mbgr +  \mbres   }_{\lp{2}}^{2}, \quad  \mbox{s.t.}  \sqb{\mbres}_{n} = \sum\limits_{m=0}^{\Mr-1} \cm \delta\sqb{n - \nm}  \notag \\
&\sqb{ \mbgr }_{n} = { \sum\limits_{k = 0}^{K - 1} {\gammar \sqb{k}\bfr  \rob{{n - \frac{\taur\sqb{k}}{T}} } }  }  , \ \cm\in2\lambda\Z
\end{align}
where $\{\mgammar,\mtaur\}$ and $\{\mcm,\mnm\}$ are the vector form of $\{\gammar,\taur\}$ and $\{\cm,\nm\}$, respectively. Despite the non-trivial setup of the additive \SR problem \eqref{eq:global problem}, the difference between signal subspaces spanned by $\fr\rob{t}$ and $u\rob{t}$ inspires us to separate the distinct signal components via sieving strategy. Namely, \eqref{eq:global problem} can be split into two tractable sub-problems: $\PO$ tackles the recovery $\mbres$ (residue) via sparse approximation on the continuum and $\PT$ solves the high-resolution spike estimation of $\sr$ via sparse deconvolution. 

\subsection{Sub-Problem $\PO$: Residue Recovery.} Assume that $\{\mgammar,\mtaur\}$ is known, \eqref{eq:global problem} amounts to,
\begin{equation}
\label{eq:PO}
\pbox{P}{1}  \quad  
 \mathop {\min}
_{\substack{
\mcm,
\mnm
}}
\     \norm{  \mbtres  - \mbres  }_{\lp{2}}^{2}, \ \mbox{s.t.} \;\; \mbtres = \mbgr - \mbyr     
, \quad \sqb{\mbres}_{n} = \sum\limits_{m=0}^{\Mr-1} \cm \delta\sqb{n - \nm}, \; \cm\in2\lambda\Z.
\end{equation}
To find the accurate solution to \eqref{eq:PO}, we leverage the continuous-time parametrization of spikes in \cite{Guo:2023:C}:
\begin{equation}
\bres[n]=\frac{\PMs ( \zm{N-1}{n} )} {\QMs ( \zm{N-1}{n})},  \quad 
\begin{array}{*{20}{l}}
\PMs \in P_{\Mr-1} \\ 
\QMs \in P_{\Mr}
\end{array}, \ \  
\zm{N-1}{n}=e^{\jmath \frac{2\pi n}{N-1}}
\label{eq:poly}
\end{equation}
where the roots of the denominator $\QMs$ are given by,
\begin{equation}
\label{eq:roots}
\QMs\rob{\um} = 0 \ \Longleftrightarrow \ \um = e^{\jmath \frac{2\pi\nm}{N-1}}, \ \ m\in\id{\Mr}.
\end{equation}
In the presence of data distortion, the model on the continuum \eqref{eq:poly} allows for refinements on residue recovery within the alternating minimization scheme.  
Hence, \eqref{eq:PO} translates to,
\begin{equation}
\label{eq:frest}
\mathop {\min}\limits_{{\PMs},{\QMs}} \  \sum\limits_{n \in\id{N-1}} {{{\left|   \frac{\btres[n] \QMs ( \zm{N-1}{n}) - \PMs ( \zm{N-1}{n} )} {\QMs ( \zm{N-1}{n})} \right|}^2}}.
\end{equation}

\bpara{Algorithmic Implementation.} Solving \eqref{eq:frest} is still challenging due to its rational polynomial structure \cite{Steiglitz:1965:J}. To address this issue, we utilize an iterative minimization strategy by assuming $\QMs^{[j-1]} \approx \QMs$, where \eqref{eq:frest} thereby reduces to,
\begin{equation}
\label{eq:linear_fitting}
\begin{Bmatrix*}[c]
\PMs^{[j]} \\ 
\QMs^{[j]}
\end{Bmatrix*} 
= \mathop{\rm arg\,min}\limits_{\{{\PMs},{\QMs}\}  }   \sum\limits_{n \in\id{N-1}} {{{\left|   \frac{\btres[n] \QMs ( \zm{N-1}{n}) - \PMs ( \zm{N-1}{n} )} {\QMs^{[j-1]} ( \zm{N-1}{n})} \right|}^2}}.
\end{equation}
It can be seen that, \eqref{eq:linear_fitting} is linear about $\PMs$ and hence, can be considered to be minimized on $\QMs$ alone. To find $\QMs$, we initialize $\QMs^{[0]}$ and acquire a collection of estimates for $\QMs$ by iteratively minimizing \eqref{eq:linear_fitting}. Among the estimates, we choose the one that minimizes the mean-squared error in \eqref{eq:linear_fitting} on $\mbtres$. Moreover, different initialization seeds for $\QMs^{[0]}$ allow for diverse estimates and thus stabilize the spike estimation. 

Once $\QMs$ is retrieved, $\PMs$ can be obtained via least-squares and the residue parameters can be computed by evaluating the residue (in complex analysis sense) of $\PMs\rob{z}/\QMs\rob{z}$ as,
\begin{equation}
\label{eq:parameter estimates}
\left\{
\begin{array}{*{20}{l}}
\cm   = \tfrac{ -(N-1)\um^{-1}\PMs\rob{\um}  }{  \rob{1 - \um^{-\rob{N-1}}} {\left. { \partial_{z} \QMs\rob{z}} \right|_{z = \um}}  }   \\
\nm  =\tfrac{(N-1)\mathsf{Im} \rob{ \log \um}}{2\pi}
\end{array}\right.
,\ \um \EQc{eq:roots} e^{\jmath \frac{2\pi\nm}{N-1}}.
\end{equation}

Next, we provide an algorithmic implementation for solving \eqref{eq:linear_fitting}: the polynomials $\{\PMs(\zm{N-1}{n}), \QMs(\zm{N-1}{n})\}_{n\in \id{N-1}}$ in \eqref{eq:linear_fitting} can be written in matrix form as,
\[
\bigl[\PMs(\zm{N-1}{n})\bigr] = \mathbf{W}_{N-1}^{\Mr} \pMs
\ \mbox{ and } \
\bigl[\QMs(\zm{N-1}{n})\bigr] = \mathbf{W}_{N-1}^{\Mr + 1} \qMs
\]
where $\{\pMs,\qMs\}$ are the coefficients of $\{\PMs \in P_{\Mr-1}, \QMs \in P_{\Mr}\}$. With the estimate $\qMs^{[j]}$ known, the minimization at $j+1$-iteration can be characterized algebraically as,
\begin{equation}
\label{eq:vector frequency matrix form}
\{  \pMs^{[j+1]},\qMs^{[j+1]}\}=\mathop{\rm arg\,min}\limits_{\{\pMs, \qMs\}}\left\|\Aj \qMs-\Bj \pMs \right\|^{2}_2
\end{equation}
where $\{\Aj,\Bj\}$ are respectively given by, 
\begin{align}
\label{eq:matrix}
\Aj = \dk( \mbtres) \Rj \mathbf{W}_{N-1}^{\Mr+1}, 
\quad 
\Bj = \Rj \mathbf{W}_{N-1}^{\Mr}, \quad 
\Rj = \left(\dk \left( \mathbf{W}_{N-1}^{\Mr + 1} \qMs^{[j]} \right)\right)^{-1} .
\end{align}
Moreover, note that, the set of minimizers is closed under scalar multiplication, meaning that $\forall \gamma \neq 0$, $\gamma \qMs$ is the minimizer to \eqref{eq:linear_fitting} if $\qMs$ minimizes \eqref{eq:linear_fitting}. To ensure the uniqueness of the estimates to $\QMs$, we use a linear normalization constraint given by 
$
\inner{\qMs^{[0]}}{\qMs^{[j+1]}} = 1
$
where $\qMs^{[0]}$ is the initialization seed for the iterative minimization algorithm. Consequently, the linearly constrained quadratic minimization \eqref{eq:linear_fitting} can be formulated as,
\begin{align}
\label{eq:opt} 
\begin{Bmatrix*}[c]
\pMs^{[j+1]}\\	
\qMs^{[j+1]}
\end{Bmatrix*} =\mathop{\rm arg\,min}\limits_{\{\pMs, \qMs\}}\left\| \Gj \zj   \right\|^{2} , \; \mbox{s.t.} \;  \bcoef^{\Hermit} \zj = 1 , \ \Gj = \begin{bmatrix}
\Aj \;\; -\Bj
\end{bmatrix}, \; 
\zj = \begin{bmatrix}
\qMs\\
\pMs
\end{bmatrix}, \;
\bcoef = \begin{bmatrix}
\qMs^{[0]}\\
\mathbf{0}
\end{bmatrix}.
\end{align}
\eqref{eq:opt} is convex and can be solved efficiently with closed-form, explicit solution. 
Eventually, to ensure the on-grid constraint $\cm\in2\lambda\Z$, we apply the quantization operator on the estimate
\begin{equation}
\label{eq:QO}
\bres[n]=\QO{\frac{\PMs ( \zm{N-1}{n} )} {\QMs ( \zm{N-1}{n})}}, \quad  \ \QO{\cdot}=2\lambda \flr{\frac{{(\cdot) +\lambda}}{{2\lambda}}}.
\end{equation}

\subsection{Sub-Problem $\PT$: Spike Estimation.} With $\mbres$ estimated from $\PO$, \eqref{eq:global problem} amounts to,
\begin{equation}
\label{eq:PT}
\pbox{P}{2} \quad \quad  
\mathop {\min}
_{\substack{
\mgammar,\mtaur
}}
\     \norm{  \mbtgr  - \mbgr  }_{\lp{2}}^{2}, \quad \mbox{s.t.} \;\; \mbtgr = \mbres + \mbyr   , \quad 
\sqb{ \mbgr }_{n} = { \sum\limits_{k = 0}^{K - 1} {\gammar \sqb{k}\bfr  \rob{{n - \frac{\taur\sqb{k}}{T}} } }} 
\end{equation}
which is a typical \SR problem that has been widely studied in different communities \cite{Candes:2013:J,Catala:2019:J,Denoyelle:2019:J,Wang:2022:J,Guo:2025:J,DeFigueiredo:1982:J,Blu:2008:J,Seelamantula:2014:J,Bhandari:2016:J}.
Despite the state-of-the-art \SR techniques, \eg $\lp{1}$-norm minimization \cite{Wang:2022:J} and non-convex optimization \cite{Kuo:2020:J}, in this paper, we relax and convert \eqref{eq:PT} into spectral fitting problem. Our solution strategy offers two key advantages: 
\begin{enumerate}
    \item {\bf Data Volume.} In the context of high-dimensional imaging, the measurements are usually sampled with fine temporal resolution, yielding large data volume. This causes stability and scalability issues for the mainstream optimization-based techniques \cite{Wang:2022:J,Kuo:2020:J,Perrone:2016:J}. 

    \item {\bf Digital Resolution.} 
Leveraging digital \SR via \usf, the sieving method provides a relatively accurate signal estimate for $\mbgr$, thereby reducing the computational burden of \SR spike estimation. In this context, even a simple signal deconvolution method achieves accurate spike estimation, as validated through extensive numerical and hardware experiments (see \secref{sec:sim} and \secref{sec:HW})
\footnote{While incorporating advanced \SR techniques \cite{Guo:2025:J,Catala:2019:J,Denoyelle:2019:J} specifically tailored for off-the-grid \usf imaging could further enhance performance in terms of resolution and speed, such efforts fall beyond the scope of this work and are left for future research.}.
\end{enumerate}

\begin{algorithm}[!t]
\setstretch{1}
\small
\caption{Robust Off-the-Grid \usf Recovery via \sritersis Algorithm.}
\begin{multicols}{2}
\begin{algorithmic}[1]
\label{alg:1}
\REQUIRE $\mbyr$, $K$ and $\Mr$.
\STATE Initialize $\mbtres^{[0]} \leftarrow \mbyr$.
\FOR{{$i = 1$} to $i_{\rm max}$}
\STATE Initialize $\MSE \leftarrow \infty$.
\FOR{\texttt{loop = 1} to \texttt{max. initializations}}
\STATE Initialize $\qMs$ as $\qMs^{[0]}$.
\FOR{$j=1$ to $j_{\rm max}$}
\STATE Construct the matrices in~\eqref{eq:matrix}.
\STATE Update $\{\qMs^{[j]}, \pMs^{[j]}\}$ by solving~\eqref{eq:opt}.
\STATE Reconstruct $\mbres^{[j]}$ via \eqref{eq:poly} and \eqref{eq:QO}.
\STATE Compute:\\ $\MSE^{[j]} = {\| \mbtres^{[i-1]} - \mbres^{[j]} \|}_{\lp{2}}^{2}/(N-1)$.
\IF{$\MSE^{[j]} < \MSE$ holds}
\STATE Update $\{ \pMs^{[i]}, \qMs^{[i]}\} \leftarrow \{\pMs^{[j]}, \qMs^{[j]}\}$. 
\ENDIF
\ENDFOR
\ENDFOR
\STATE Compute $\mbtres^{[i]}$ with $\{ \pMs^{[i]}, \qMs^{[i]}\}$ via \eqref{eq:poly} and \eqref{eq:QO}.
\STATE Compute $\htsr$ by solving \eqref{eq:nPT}.
\STATE Estimate $\htsr\sqb{0}$ using matrix pencil for $\{\htsr\sqb{l}\}_{l=1}^{\Ifr}$.
\STATE Update $\htsr\sqb{0}$ via \eqref{eq:c0}.
\STATE Update $\{\mtgammar^{[i]},\mttaur^{[i]}\}$ by solving \eqref{eq:nPT} with matrix pencil.
\STATE Recover $\mbtgr^{[i]}$ with $\{\mtgammar^{[i]},\mttaur^{[i]}\}$ via \eqref{eq:g}.
\IF{$\| { \mbtgr^{[i]} - \mbtgr^{[i-1]}  }\|_{\lp{\infty}} \leqslant \sigma$ holds} 
\STATE Update $\{ \tpMs, \tqMs, \mtgammar,\mttaur\} \leftarrow \{ \pMs^{[i]}, \qMs^{[i]}, \mtgammar^{[i]},\mttaur^{[i]} \}$. 
\STATE {Terminate all loops}.
\ENDIF
\STATE Update $\mbtres^{[i]}$ as $\mbtres^{[i]} \leftarrow  \mbtgr^{[i]} - \mbyr  $.
\ENDFOR
\STATE Reconstruct $\tsr\rob{t}$ via \eqref{eq:SRF}.
\ENSURE $\{ \tpMs, \tqMs, \mtgammar,\mttaur\}$ and $\tsr\rob{t}$. 
\end{algorithmic}
\end{multicols}
\end{algorithm}

Using the analytical representation \eqref{eq:bgr}, we relax \eqref{eq:PT} as,
\begin{equation}
\label{eq:nPT}
(\mbox{Relaxed}\; \pbox{P}{2}) \quad   
\mathop {\min}
_{\substack{
\mgammar,\mtaur
}}
\     \sum\limits_{\abs{l}\leqslant \Ifr} \abs{   \htsr \sqb{l} - \sum\limits_{k \in\id{K}} \gammar \sqb{k}  e^{\jmath \tfrac{- 2l \pi  \taur\sqb{k}}{\tau}}   }^{2} , \; \;  \mbox{where} \ \  \htsr\sqb{l} = \frac{\htgr \sqb{l}}{\hfr\sqb{l}} , \;\ \ \tgr = \Delta^{-1} \rob{\bres + \byr  }
\end{equation}
where $\Delta^{-1}(\cdot)$ denotes the anti-difference operator. The relaxed sub-problem \eqref{eq:nPT} can be solved efficiently using high-resolution spectral estimation approaches \cite{Stoica:1989:J,Hua:1990:J,Bhaskar:2013:J,Park:2018:J,Guo:2022:J}. 

\bpara{Algorithmic Implementation.} From \eqref{eq:nPT}, $\tgr$ can be recovered up to an unknown constant, resulting in the indeterminacy of $\htgr\sqb{0}$ as well as $\htsr\sqb{0}$. To address this issue, we abstract the problem as follows: let $\htsr\sqb{l} = \hsr\sqb{l} + \cz \delta\sqb{l}$, where $\hsr\sqb{l}= \sum_{k \in\id{K}} \gammar \sqb{k}  e^{\jmath {- 2l \pi  \taur\sqb{k}}/{\tau}}   $, $l\in\sqb{-K,K}$ and $\cz\in\R$ is an unknown constant. Then, from \eqref{eq:AF}, we know that,
\begin{equation}
\label{eq:eig}
\G{\htsr} = \G{\hsr} + \cz \Id \; \Longrightarrow \; \G{\htsr} \mf = \cz  \mf
\end{equation}
which indicates that $\cz$ is the eigenvalue of $\G{\htsr}$. Moreover, the structure of $\sr\rob{t}$ and Toeplitz matrix results in,
\begin{equation}
\notag
\left\{
\begin{array}{*{15}{l}}
\gammar\sqb{k} \in\R, k\in\id{K}\\
\taur\sqb{k} \neq \taur\sqb{k'}, \forall k\neq k'
\end{array}\right.
\ 
\Longrightarrow
\ 
\left\{
\begin{array}{*{15}{l}}
\G{\hsr} = (\G{\hsr})^{\Hermit} \\
\rank\rob{\G{\hsr} } = K
\end{array}
\right.
\end{equation}
which suggests an algebraic solution for estimating $\cz$:
\begin{equation}
\label{eq:c0}
\htsr\sqb{0} \approxeq \hsr\sqb{0}
\; \Longrightarrow \;
\begin{array}{*{15}{c}}
\abs{\cz} = \mathop {\min}
_{\substack{
\mf
}} \ \norm{ \G{\htsr} \mf }_{\lp{2}}^{2} \\
\mbox{subject to} \; \norm{\mf}_{\lp{2}} = 1	
\end{array}.
\end{equation}
With $\Ifr > K$, we can obtain an initial estimate to $\hsr\sqb{0}$ by first performing spectral fitting on $\{\htsr\sqb{l}\}_{l=1}^{\Ifr}$ and then estimate $\hsr\sqb{0}$ with the retrieved spectral parameters $\{\tgammar\sqb{k},\ttaur\sqb{k}\}_{k\in\id{K}}$.
This empirically results in a small value of $\abs{\cz}$ and hence \eqref{eq:c0} holds.
Since $\cz\in\R$, we can determine its sign by evaluating the eigendecomposition of $\G{\htsr}$ \eqref{eq:eig}. 
Having $\{\htsr\sqb{l}\}_{\abs{l}\leqslant \Ifr}$ known, we solve \eqref{eq:nPT} using the matrix pencil method \cite{Hua:1990:J}\footnote{Other state-of-the-art high-resolution spectral estimation techniques such as atomic-norm methods \cite{Bhaskar:2013:J,Park:2018:J} are also applicable in our context.} and reconstruct $\mbgr$ via \eqref{eq:g} and apply the finite difference operator. 
We summarize the procedure of \sritersis in \algref{alg:1}\footnote{$\sigma$ is choose based on the measurement noise level. In the context of low-resolution sampling, $\sigma \propto 2\lambda/2^{B}$ where $B$ is the quantization bit budget.}.  

\section{\ussr for ToF Imaging}
\label{sec:ToF}

To demonstrate practical impact, we apply our approach to ToF imaging (see Chap.~5, \cite{Bhandari:2022:Book} for overview), where each pixel captures a scene-dependent time profile with temporal resolution on the order of nanoseconds to picoseconds \cite{Pellegrini:2000:J,Bhandari:2015:J,Bhandari:2016:J,Heide:2013:J}. We use $\mathbf{r} = [x \ \  y]^{\top}$ to denote a point in the Cartesian coordinate where $(\cdot)^{\top}$ is the transpose operation.
Acting as active imaging systems, ToF sensors probe the \ttd scene of interest with some time-concentrated kernel \cite{Guo:2025:J}, represented by $\pr \rob{t}$ at a point $\mat{r}$. 

The spatio-temporal scene response function (SRF) $\srr \rob{t}$ characterizes the \ttd scene. In the general multi-reflection scenarios (see \fig{fig:sim_SR}-\fig{fig:HW_14}), the SRF is given by, 
\begin{equation}
\label{eq:SRFr}
\srr \rob{t} = \sum\limits_{k = 0}^{K - 1} {\gammarr \sqb{k}\delta  \rob{t - \taurr\sqb{k}} }
\end{equation}
where $\{\gammarr \sqb{k},\taurr\sqb{k}\}_{k\in\id{K}}$ are the corresponding reflectivities and time-delays ($\taur\sqb{k} = 2 {d_{\mathbf{r}}}\sqb{k}/c$, $c$ is the light speed) induced by $K$ light paths at point $\mat{r}$. 
As a result, the reflected signal is given by $\rrr\rob{t} =   \rob{\pr \ast \srr}  \rob{t}$, describing the interaction between the emitted signal $\pr$ with the \ttd scene $\srr$.

\begin{figure*}[!tb]
\centering
\resizebox{\textwidth}{!}{
$
\underbrace{\pr}_{\text{Emitted Signal}} \to \underbrace{\boxed{\boxed {\srr}}}_{\text{\srf}} \to \underbrace{\color{black!100} \rrr\rob{t} =   \rob{\pr \ast \srr}  \rob{t} }_{\text{Reflected Signal}} \to \underbrace{\boxed{\boxed\phir }}_{\text{IRF}} \to \underbrace{ \color{black!100} \grr\rob{t} = \rob{\rrr \ast \phir}  \rob{t}  }_{\text{Measured Signal}} \ \xrightarrow{{{\textsf{USF}}}} \underbrace{\yrr\rob{t} = \MO{\grr\rob{t}}}_{\text{Folded LDR Signal}}  \ \xrightarrow{{{\textsf{Sampling}}}} \ \underbrace{\yrr \sqb{n} = {\left. {\yrr \left( t \right)} \right|_{t = nT}}}_{\text{Measured Samples}}
$}
\caption{Block diagram for modulo ToF image formation process. The goal is to estimate $\srr\rob{t}$ from $\{\yrr \sqb{n}\}_{n\in\id{N}}$. }
\label{fig:MToF}
\end{figure*}

The reflected signal is captured at the ToF sensor through its electro-optical architecture, which is characterized by its instrument response function (IRF), denoted by $\phir\rob{t}$. Consequently, the continuous-time measurements read,
$
\grr\rob{t} = \rob{\rrr \ast \phir}  \rob{t},
$
which can be further simplified as
\begin{equation}
\label{eq:ToF measurements}
\grr\left( t \right) = \left( \srr \ast \fr \right) (t)\quad  \mbox{and} \quad \fr (t) \DE  \rob{\pr  \ast \phir} (t)
\end{equation}
where $\fr$ represents the kernel characterized in \eqref{eq:BL}. Plugging \eqref{eq:SRFr} into \eqref{eq:ToF measurements}, $\grr$ can be re-expressed as,
\begin{equation}
\label{eq:gr}
\grr\left( t \right) \EQc{eq:ToF measurements} \sum\limits_{k = 0}^{K - 1} {\gammarr \sqb{k}\fr  \rob{{t - \taurr\sqb{k}} } }
\end{equation}
In the context of multiple reflections, the mixture of far and close targets results in \emph{weak-strong} characteristics, as shown in \fig{fig:WS} and \fig{fig:3Bp50}. This requires concurrent \HDR and digital \SR capabilities of the imaging pipeline, which is impractical with conventional approaches in practice, since:
\begin{enumerate}[leftmargin=20pt, label = \uline{\arabic*}),labelsep = 2pt]
\item \textbf{Hardware Cost.} The circuitry complexity increases largely with the quantization budget. 
\item \textbf{Power Consumption.} The power consumption of the ADC scales exponentially with the quantization budget \cite{Walden:1999:J}.
\item \textbf{Data Volume.} Since ToF sensors capture data at high temporal resolution, the data volume will explode with higher quantization resolution. 
\end{enumerate}
Conventional DR–DRes trade-off limit practical \SR in ToF imaging. We show that \usf overcomes this, enabling temporally super-resolved ToF recovery from low-resolution data with consistent $\geqslant 30$ dB gains over traditional methods. A global view of the \usf-ToF image formation is shown in \fig{fig:MToF}. For continuous-wave ToF imaging within the USF, we refer the reader to \cite{Shtendel:2022:C}.

\section{Numerical Experiments}
\label{sec:sim}

HDRes is essential for achieving temporal \SR from low-resolution measurements---an ability lacking in conventional methods \cite{Wang:2022:J,Candes:2013:J,Catala:2019:J,Chi:2020:J}. All experiments in this study are based on ToF hardware measurements. The distinction between numerical and hardware experiments lies in whether the hardware-acquired data are re-digitized using the \madc or directly processed in their original form. To showcase the performance gain of the \usf, we present a series of numerical experiments, including:
\begin{enumerate}[leftmargin=30pt, label = \uline{\arabic*})]
\item {\bf Clipping-free recovery:} \HDR recovery when the conventional ADC clips.
\item {\bf Low-resolution sampling:} ToF Imaging using few quantization budget.
\item {\bf \SR Imaging:} Close inter-object separation on semi-real ToF measurements.
\item {\bf High-order Imaging}: Concurrent weak-strong targets detection on semi-real ToF measurements.
\end{enumerate}
The experimental parameters including sampling step $T$ and bit budget $B$, are tabulated in \tabref{tab:sim}. The peak-signal-to-noise ratio (PSNR) and $\norm{\gr}_{\Lp{\infty}}$ are utilized to evaluate reconstruction quality and DR extension, respectively. 

\begin{table}[!t]
\centering
\caption{Numerical Experiments: Parameters and Performance Metrics.}
\label{tab:sim}
\resizebox{0.7\textwidth}{!}{
\begin{tabular}{@{}lccccccccccc@{}}
\toprule
\multicolumn{1}{c}{Figure}   & Bits &$K$ &$T$ & $M$ & $\norm{\g}_{\Lp{\infty}}$ & $\lambda$  &$ \tk$ & $\psnr$
\\ \midrule
&                          &     &               (ns)            &    &  (V) &   (V) &  (ns)  &  (\dB)           \\ \midrule
\multicolumn{1}{c}{\fig{fig:clipping}}&$4$&$2$&$0.48$&$108$&$0.99$&$0.05$&$[30.38,42.07]$&$39.83$\\
\multicolumn{1}{c}{\subfig{fig:sim_SR}{a}{}}&$3$&$2$&$0.77$&$12$&$0.98$&$0.10$&$[30.40,42.14]$&$39.05$\\
\multicolumn{1}{c}{\subfig{fig:sim_SR}{b}{}}&$3$&$2$&$0.77$&$12$&$1.00$&$0.10$&$[32.27,42.13]$&$39.81$\\
\multicolumn{1}{c}{\subfig{fig:sim_SR}{c}{}}&$3$&$2$&$0.77$&$12$&$0.99$&$0.10$&$[34.23,42.17]$&$40.24$\\
\multicolumn{1}{c}{\subfig{fig:sim_SR}{d}{}}&$3$&$2$&$0.77$&$12$&$0.99$&$0.10$&$[36.73,42.54]$&$41.10$\\
\multicolumn{1}{c}{\fig{fig:3Bp50}}&$3$&$3$&$0.77$&$44$&$0.99$&$0.10$&$[84.97,97.19,111.04]$&$40.68$\\
\multicolumn{1}{c}{\textemdash}&$3$&$3$&$0.77$&$38$&$1.00$&$0.10$&$[84.80,97.11,111.09]$&$36.34$\\
\bottomrule
\end{tabular}
}
\end{table}

\begin{SCfigure}
\centering
\includegraphics[width=0.7\linewidth]{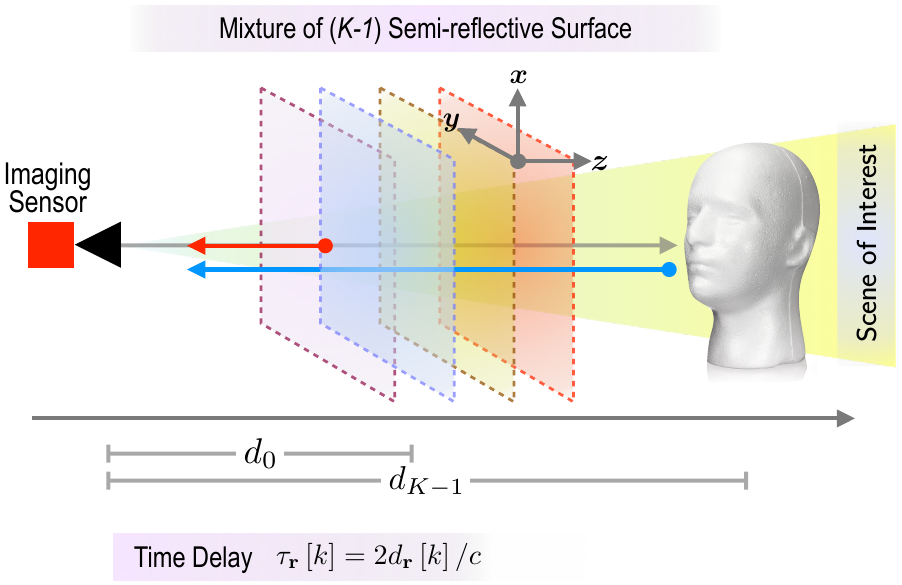}
\caption{Experimental setup for \SR ToF Imaging. The \ttd scene comprises of a mannequin head positioned between a diffusive semi-translucent surface and a wall in the backdrop. The diffusive sheet is moving closer to the mannequin head, leading to the challenges of separating two close objects. }
\label{fig:Scene}
\end{SCfigure}

\subsection{Performance Evaluation with Conventional Approaches.}
\label{subsec:HDR}

\subsubsection{Clipping-free \HDR Recovery.} 

In the first experiment, we investigate the \HDR imaging case where $\norm{g}_{\Lp{\infty}} = 20\lambda$. The waveform $\gr$ arises from hardware experiments \cite{Bhandari:2020:J,Guo:2025:J} where the \ttd scene comprises of a mannequin head placed between a diffusive semi-translucent surface and a wall in the backdrop, as shown in \fig{fig:Scene}. The inter-object separation between the diffusive surface and mannequin head is $1.6$ m. The raw data comprising of $120\times 120\times 794\times 4$ image tensor is obtained from a lock-in ToF sensor, where $N=794$ refers to the number of ToF measurements recorded with sampling period $T=480.75$ ps. We use a single-pixel data ($\mathbf{r}=[100\;50]^{\transp}$) and acquire $\{\yrr\sqb{n}\}$ via \eqref{eq:map}. 

\begin{SCfigure}[]
\centering
\includegraphics[width=0.7\linewidth]{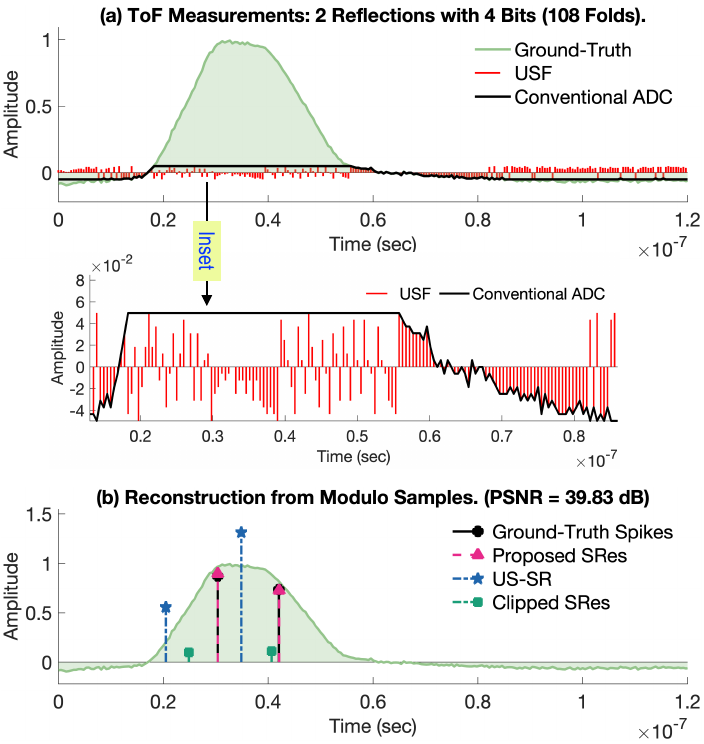}
\caption{Clipping-free \HDR recovery ($\norm{g}_{\Lp{\infty}} = 20\lambda$). On the acquisition front, the DR of the conventional ADC is the same as the \madc, where the resulting data clipping causes inaccurate time-delay estimation. On the algorithm front, the \USSR approach suffers from spectral leakage and dense folds, leading to large deviation in \ttd scene recovery. In contrast, the proposed method offers accurate signal recovery, where the estimated time-delays are visually indistinguishable from the ground-truth.	
}
\label{fig:clipping}
\end{SCfigure}

To highlight the benefits of paradigm shift from conventional acquisition to \usf, 
we set the DR of the conventional ADC to $2\lambda$, the same as \madc. This yields data clipping as shown in \subfig{fig:clipping}{a}{}. To benchmark the performance of our approach, we compare the reconstructions obtained from: i) conventional ADC measurements $+$ \SR technique for solving \eqref{eq:nPT} and ii) \madc measurements $+$ the state-of-the-art \USSR method \cite{Bhandari:2022:J}. 

As shown in \subfig{fig:clipping}{b}{}, the conventional ADC data suffers from clippling and results in failure of \srf recovery. The \USSR has difficulty in finding dense spikes, causing erratic time-delay estimation. Despite the \HDR challenge, our approach offers accurate signal recovery with $\psnr = 39.83$ \dB, and the time-delay estimation is visually indistinguishable from the ground-truth.  
This result pinpoints the advantages of the \sritersis approach at both acquisition and algorithmic recovery fronts.

\subsubsection{Low-Resolution ToF Imaging.} 

Quantization bit budget is a key constraint in high-bandwidth ToF sensing, where power consumption and data volume grow rapidly with bit resolution \cite{Walden:1999:J,Kadambi:2013:J}. Existing low-resolution approaches, such as one-bit sampling \cite{Bhandari:2020:J}, rely on prior DR knowledge and still suffer from clipping and saturation (see \fig{fig:clipping}), limiting digital super-resolution for close–far scene recovery (see \fig{fig:WS}). Our proposed method overcomes these limitations, enabling super-resolved scene recovery from low-resolution measurements. We demonstrate this using weak–strong reflections: $\gr\sqb{n} = \Gamma\sqb{1}\fr\rob{nT-\taur\sqb{1}}+\Gamma\sqb{2}\fr\rob{nT-\taur\sqb{2}}$ with $\abs{\Gamma\sqb{1}/\Gamma\sqb{2}}=10$, $\abs{\taur\sqb{1}-\taur\sqb{2}}=75$, and $N=501$. Conventional ADC and \madc data are quantized with identical bit budgets ($B=3$–15) while varying DR ($\norm{g}_{\Lp{\infty}}={10,20,30}\lambda$). Mean-squared error (MSE) is computed over $5000$ randomized trials to produce the performance curves shown in \fig{fig:curve}.

The main conclusions from \fig{fig:curve} are:
\begin{enumerate}[leftmargin=45pt, label = \pbox{C}{\arabic*:}]
\item {\bf Consistent Performance Gain.} Our method provides a precision improvement (at least $30$~dB) across all DR extensions and quantization resolutions, even in low-resolution regimes ($B \leqslant 6$).
\item {\bf Robustness to Input DR.} Our method remains insensitive to input DR, offering stable performance even with a 3x boost in the input DR.
\end{enumerate}

\subsection{Super-Resolved Imaging on Semi-Real ToF Data.}
\label{subsec:SR}

\begin{figure}[!t]
 \centering
  \captionsetup{width=.85\linewidth}
\includegraphics[width=\linewidth]{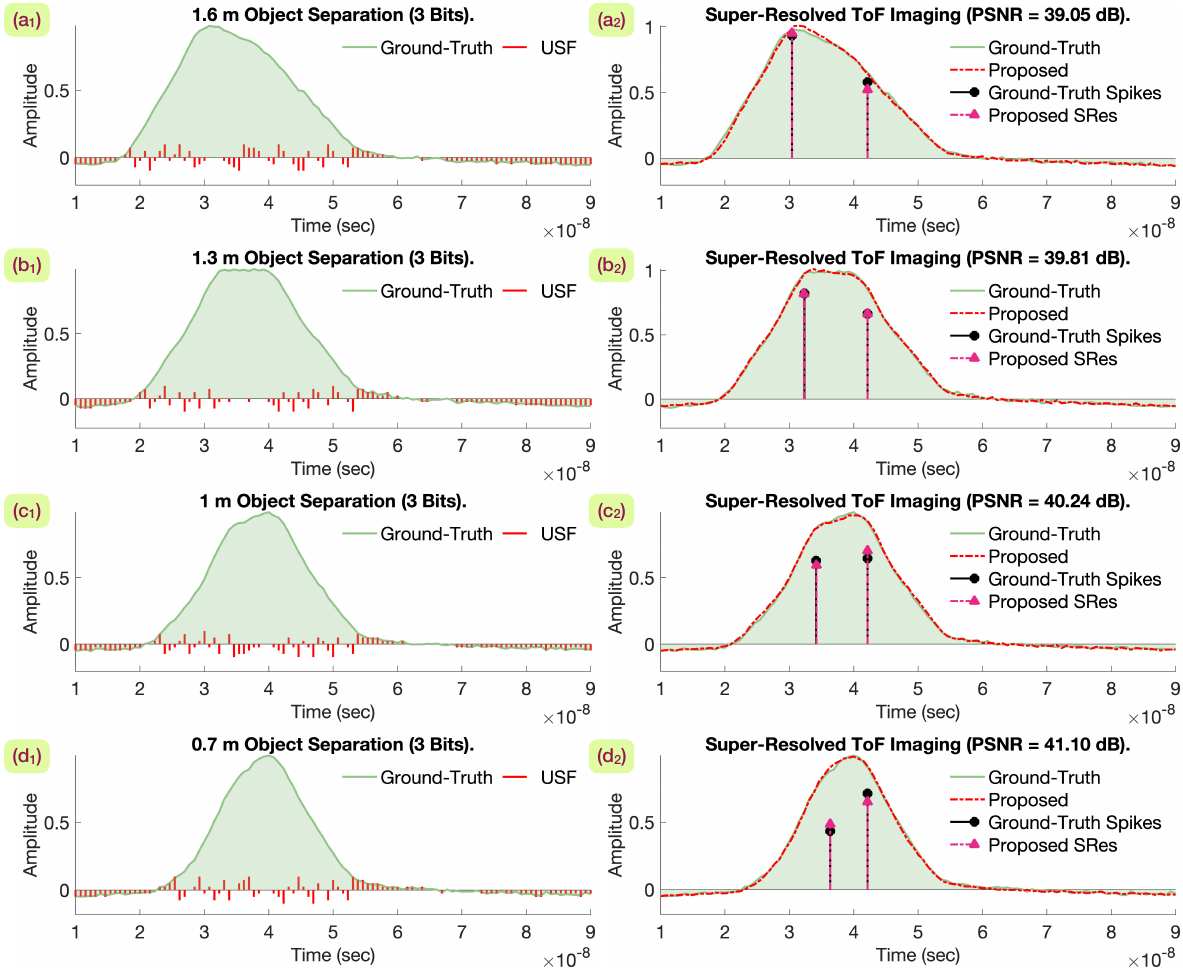}
\caption{\SR ToF imaging. The \ttd scene consists of a mannequin head placed between a diffusive translucent surface and a wall (backdrop), as shown in \fig{fig:Scene}. By moving the diffusive surface, the inter-object separation is uniformly reducing from (a) $1.6$ m, (b) $1.3$ m, (c) $1.0$ m to (d) $0.7$ m, respectively. Using $3$-bits quantization, our method achieves $\norm{g}_{\Lp{\infty}}=10\lambda$ and super-resolves the inter-object separation with estimation error down to $0.9$ cm. }
\label{fig:sim_SR}
\end{figure}

In this experiment, we demonstrate that the \HDR and digital \SR of the \usf method translate into temporal \SR in inter-object separation. We use the same experimental setup as depicted in \secref{subsec:HDR} and \fig{fig:Scene}, and progressively reduce the inter-object distance between the diffusive surface and mannequin head from $1.6$ m, $1.3$ m, $1.0$ m to $0.7$ m, respectively, with $0.3$ m reduction in each experiment. This equivalently leads to equidistant spike shifts as shown in \fig{fig:sim_SR}. For this dataset, raw data comprising of $120\times 120\times 361\times 4$ image tensor is acquired from a lock-in ToF sensor, with $3$-bits quantization and sampling period $T=0.77$ ns. This gives rise to approximately the same power consumption as the configuration specified in \cite{Bhandari:2016:J} where measurements are sampled with one-bit quantization and $96.15$ ps.

\begin{figure}[!t]
 \centering
  \captionsetup{width=.85\linewidth}
\includegraphics[width=0.9\linewidth]{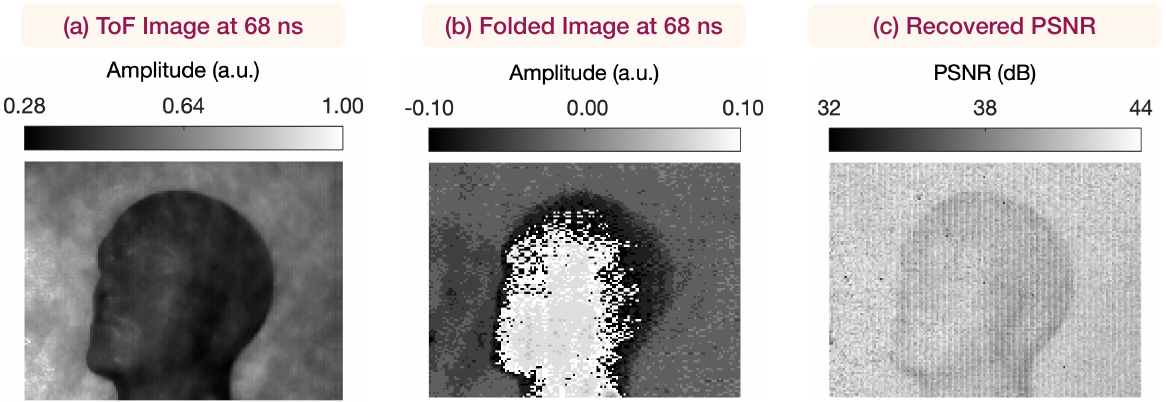}
\caption{\SR recovery of ToF measurements. (a) A slice of ToF tensor at $t=68$ ns. (b) The folded version of (a). (c) Recovered PSNR of each pixel data. Despite the low-resolution sampling settings, our method achieves accurate recovery across all pixels with $\psnr \geqslant 33$ \dB.
}
\label{fig:psnr}
\end{figure}

\begin{figure}[!t]
 \centering
  \captionsetup{width=.85\linewidth}
\includegraphics[width=\linewidth]{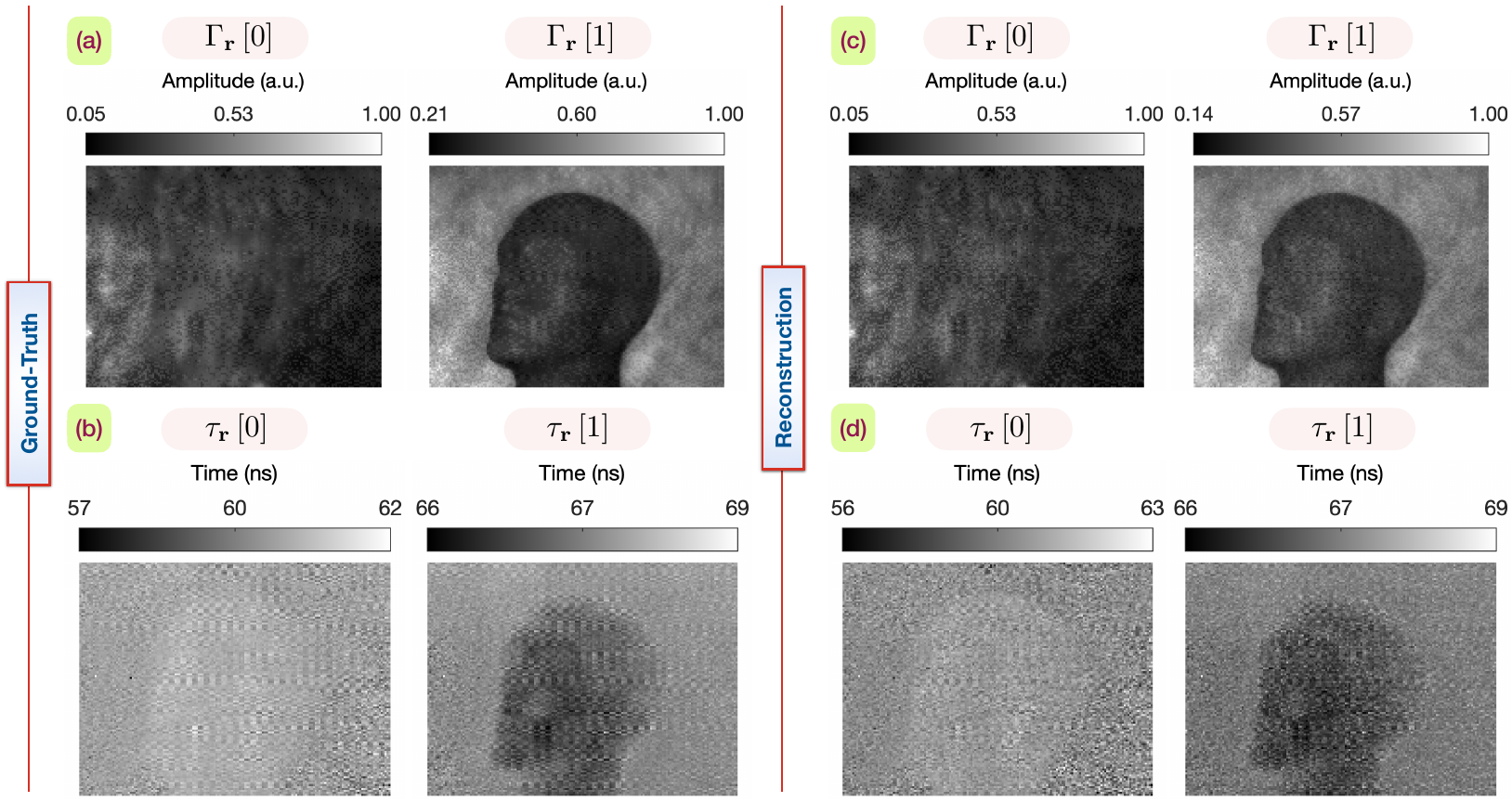}
\caption{
Super-resolving object through a diffusive semi-translucent surface. (a) and (b) are amplitude and depth imaging using $11$-bits resolution; (c) and (d) are corresponding results utilizing our \ussr method. The experimental setup is shown in \fig{fig:Scene} with inter-object separation of $0.7$ m. 
A slice of ToF image tensor at $t=68$ ns and its folded version are shown in \fig{fig:psnr}.
Using $3$-bits quantization, our method super-resolves two close objects with estimation error down to $0.29$ cm.  }
\label{fig:headSR}
\end{figure}

For per-pixel measurements, we set $\norm{\grr}_{\Lp{\infty}}=10\lambda$, inducing a true \HDR scenario. A ToF image slice at $t=68$ ns and its folded version are shown in \fig{fig:psnr}. We first evaluate single-pixel recovery ($\mathbf{r}=[60;60]^{\transp}, K=2$), where closely spaced objects make one spike nearly invisible due to kernel overlap (\subfig{fig:sim_SR}{c}{2}, \subfig{fig:sim_SR}{d}{2}). Despite this challenge, our method reconstructs the signal with ${\psnr\approx40}$ dB. Extending to the full scene ($0.7$ m separation), we recover accurate ToF profiles across all pixels, producing a reconstruction visually indistinguishable from the 11-bit ground truth (\fig{fig:headSR}). The inter-object separation errors are $\{1.26, 8.91, 9.25, 2.90\}\times10^{-3}$ m, equivalent to resolving $\{4.19, 29.69, 30.83, 9.66\}$ ps in the time domain. Achieving ps-scale precision with only 3-bit, $T=0.77$ ns acquisition demonstrates the super-resolution capability of our method.

\begin{SCfigure}
\centering
\includegraphics[width=0.72\linewidth]{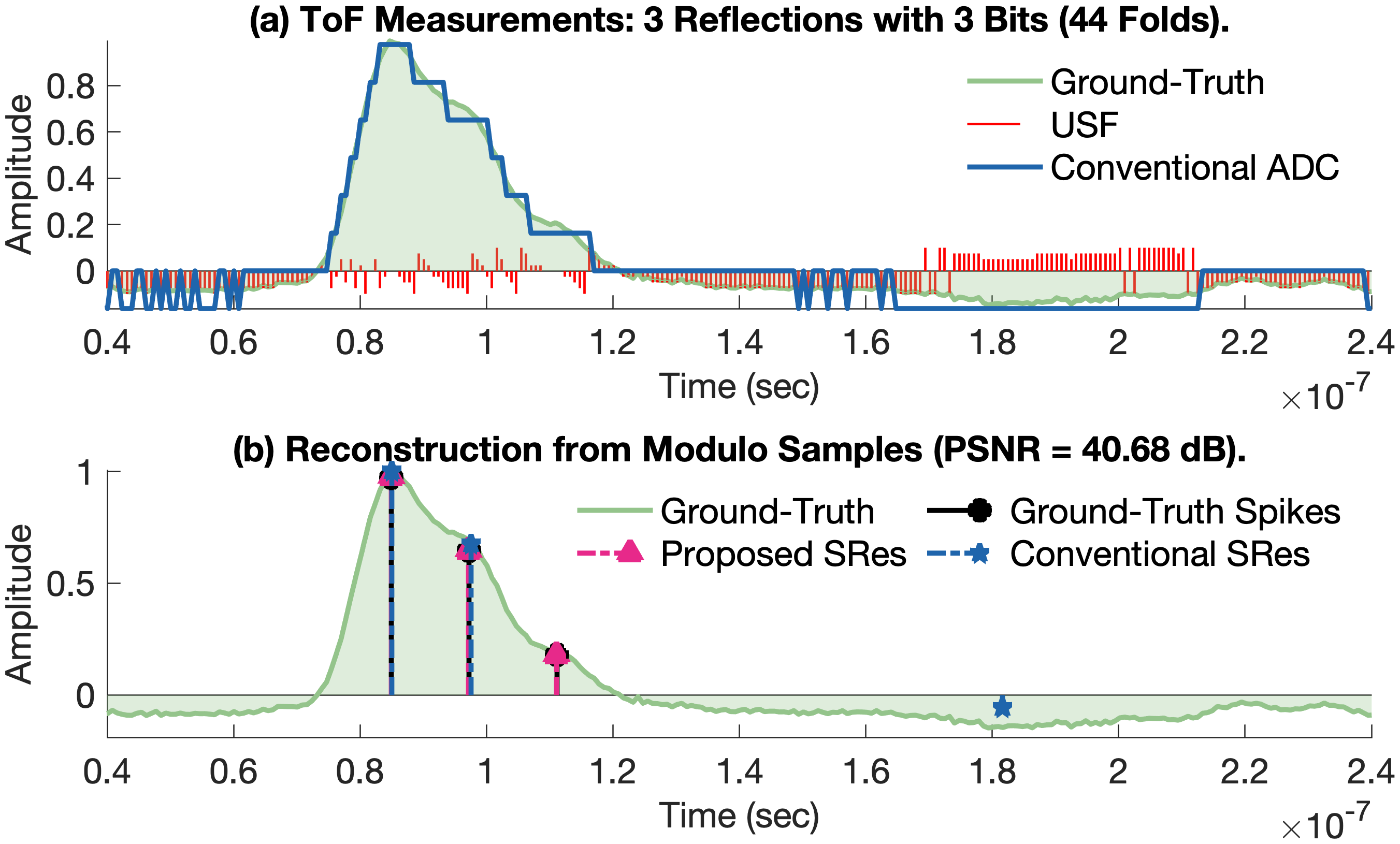}
\caption{High-order imaging of weak-strong reflections ($K=3$, $\frac{\max_{k} \abs{\Gamma\sqb{k}}}{\min_{k} \abs{\Gamma\sqb{k}}} = 5.42$). Two translucent surfaces are placed in front of a wall with separation of $1.8$ m and $2.0$ m, respectively. Using $3$-bits quantization, our method achieves $\norm{g}_{\Lp{\infty}}=10\lambda$ and  super-resolves the inter-object separation of $1.83$ m and $2.08$ m, accurately matching the experimental setup. SRes methods fail to recover the weak target due to coarse quantization resolution in conventional ADCs. }
\label{fig:3Bp50}
\end{SCfigure}

\subsection{High-Order Imaging on Semi-Real ToF Data.}
\label{subsec:High order}

This experiment is dedicated to pushing the limits of our approach in the mixed scenarios of (i) \uline{high-order multi-path imaging $K=3$} and \uline{weak-strong target detection}. We use a calibrated scene with two translucent surfaces with a wall in the backdrop. The inter-object separation is $1.8$ m and $2$ m, respectively, with sampling period of $T=770$ ps.
We evaluate the recovery of a single-pixel ToF measurements at $\mathbf{r}=[70\;50]^{\transp}$ which entails a mixture of weak-strong targets with amplitude ratio $\max_{k} \abs{\Gamma\sqb{k}}/\min_{k} \abs{\Gamma\sqb{k}} = 5.42$, as shown in \fig{fig:3Bp50}. We use $\norm{g}_{\Lp{\infty}}=10\lambda$ for modulo folding.
Using $3$-bits quantization, our approach achieves signal recovery with $\psnr = 40.68$ \dB. From the experimental results in \tabref{tab:sim}, the estimated inter-object separation is $1.83$ m and $2.08$ m, which accurately matches the experimental setup and the reported results in \cite{Bhandari:2020:J,Guo:2025:J}. Utilizing the same quantization resolution, the conventional SRes methods fail to resolve the weak object due to coarse quantization resolution, as shown in \subfig{fig:3Bp50}{b}{}. This corroborates the \SR capability and noise resilience of our method.

\section{Hardware Experiments}
\label{sec:HW}

\begin{table}[!t]
\centering
\caption{Hardware Experiments: Parameters and Performance Metrics.}
\label{tab:HW}
\resizebox{0.65\textwidth}{!}{
\begin{tabular}{@{}lccccccccccc@{}}
\toprule
\multicolumn{1}{c}{Figure}   & Bits &$K$ & $M$ & $\norm{\g}_{\Lp{\infty}}$ & $\lambda$  &$ \tk$ & $\psnr$
\\ \midrule
&                          &               &    &  (V) &   (V) &  (ns)  &  (\dB)           \\ \midrule
\multicolumn{1}{c}{\subfig{fig:HW_14}{a}{}}&$6$&$2$&$28$&$7.58$&$0.33$&$[72.67,84.91]$&$40.06$\\
\multicolumn{1}{c}{\subfig{fig:HW_14}{b}{}}&$6$&$2$&$29$&$7.76$&$0.33$&$[72.61,83.05]$&$40.78$\\
\multicolumn{1}{c}{\subfig{fig:HW_14}{c}{}}&$6$&$2$&$48$&$7.76$&$0.33$&$[72.15,79.12]$&$41.56$\\
\multicolumn{1}{c}{\textemdash}&$6$&$2$&$24$&$6.42$&$0.33$&$[72.78,85.03]$&$39.72$\\
\multicolumn{1}{c}{\textemdash}&$6$&$2$&$50$&$6.42$&$0.33$&$[72.57,83.02]$&$40.65$\\
\multicolumn{1}{c}{\textemdash}&$6$&$2$&$32$&$6.51$&$0.33$&$[71.86,79.00]$&$41.12$\\
\multicolumn{1}{c}{\textemdash}&$6$&$2$&$29$&$5.26$&$0.33$&$[72.71,84.93]$&$40.55$\\
\multicolumn{1}{c}{\textemdash}&$6$&$2$&$28$&$5.26$&$0.33$&$[72.34,82.76]$&$40.13$\\
\multicolumn{1}{c}{\textemdash}&$6$&$2$&$20$&$5.26$&$0.33$&$[72.30,79.33]$&$40.17$\\
\bottomrule
\end{tabular}
}
\end{table}

\begin{figure*}[!t]
 \centering
  \captionsetup{width=.85\linewidth}
\includegraphics[width=\linewidth]{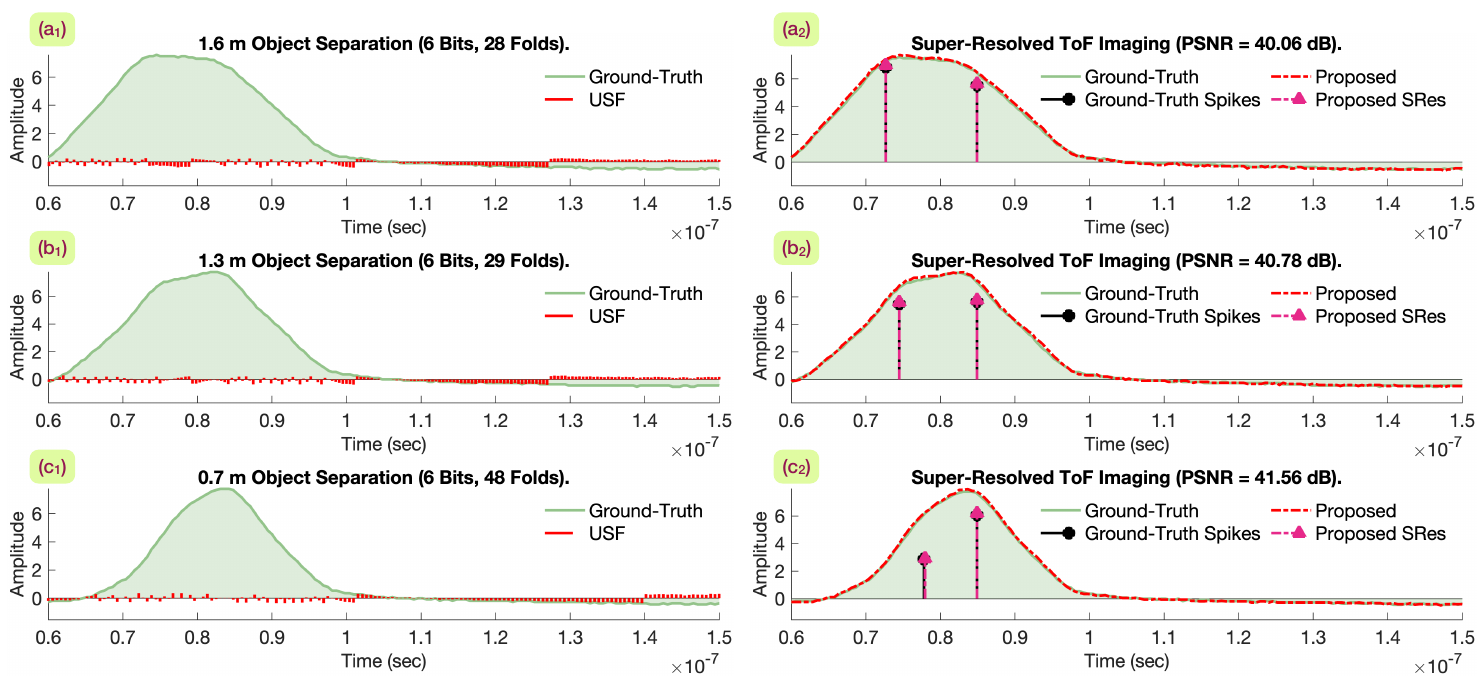}
\caption{Lock-in sensor based \SR ToF imaging. We benchmark the performance of the proposed method with experimental setup illustrated in \fig{fig:Scene} with dynamic extension $\norm{g}_{\Lp{\infty}}=23.51\lambda$. The diffusive sheet is moving closer to the mannequin head, reducing the inter-object separation from (a) $1.6$ m, (b) $1.3$ m to (d) $0.7$ m, respectively. Using $6$-bits quantization, our method super-resolves two close objects with estimation error down to $2$ cm.}
\label{fig:HW_14}
\end{figure*}

Despite progress in \SR theory and algorithms, hardware validation under low-resolution constraints remains limited. To show how digital \SR in USF enables temporal \SR in practice, we conduct ToF experiments with measurement noise and folding non-idealities \cite{Guo:2024:J,Bhandari:2021:J}. Using the setup in \fig{fig:Scene}, the distance between a diffusive surface and a mannequin head is reduced from $1.6$ m to $0.7$ m. Following \cite{Bhandari:2022:J}, ToF signals at $\mathbf{r}=[100;50]^{\transp}$ are acquired with a \madc ($\lambda=0.33$, $B=6$ bits). To stress-test performance, the DR is increased from $\norm{g}_{\Lp{\infty}}=15.94\lambda$ to $23.51\lambda$ (\fig{fig:HW_14}), with experimental parameters and results summarized in \tabref{tab:HW}.

In comparison with the experimental parameters in \cite{Guo:2025:J}, both temporal resolution and quantization resolution are relatively low, which introduces algorithmic challenges in separating two objects via SRes. Under this setup, the \USSR approach \cite{Bhandari:2022:J} fails due to measurement distortion and spectral leakage. By contrast, our \ussr method achieves significantly improved performance: the estimation error is small, even when the inter-object separation is reduced, as confirmed by experimental results:
\begin{enumerate}[leftmargin=20pt, label = \uline{\arabic*}),labelsep = 5pt]
\item $\norm{g}_{\Lp{\infty}}=15.94\lambda$: (a) $1.47\times 10^{-2}$ m, (b) $1.20\times 10^{-3}$ m and (c) $3.18\times 10^{-2}$ m . 
\item $\norm{g}_{\Lp{\infty}}=19.72\lambda$: (a) $5.09\times 10^{-3}$ m, (b) $2.42\times 10^{-3}$ m and (c) $7.33\times 10^{-2}$ m . 
\item $\norm{g}_{\Lp{\infty}}=23.51\lambda$ in \fig{fig:HW_14}: (a) $1.76\times 10^{-3}$ m, (b) $6.31\times 10^{-4}$ m and (c) $2.01\times 10^{-2}$ m . 
\end{enumerate}
This accurately matches the experimental setup and the reported results in \cite{Bhandari:2020:J,Guo:2025:J}. Despite the challenging experimental conditions, such as measurement noise and hardware imperfection, the proposed method super-resolves two objects up to a separation uncertainty of centimeter resolution, utilizing low-resolution acquisition scheme. This effectively demonstrates the \SR capability and noise resilience, which paves the way for \SR ToF imaging with low-resolution acquisition scheme \cite{Bhandari:2020:J,Bhandari:2016:J,Guo:2025:J}.

\section{Conclusion}

In this paper, we leverage the Unlimited Sensing Framework (USF) to advocate a broader notion of super-resolution---one that jointly enhances amplitude and temporal resolution. We revisit the classic problem of recovering off-the-grid spikes or Dirac impulses from filtered measurements, which is challenged in conventional systems by a trade-off between dynamic range and digital resolution. This trade-off leads to clipping of strong components or loss of weak ones beneath the quantization noise floor.

USF addresses this limitation through modulo-based nonlinear encoding at the analog front-end, enabling digital super-resolution—high measurement precision under low-bit quantization. We show that this facilitates off-the-grid sparse recovery even under extreme constraints (\eg \ 3-bit resolution), unlocking temporal super-resolution beyond conventional bounds. We provide new theoretical guarantees for non-bandlimited kernels and introduce a robust recovery algorithm that accounts for hardware imperfections. Using time-of-flight (ToF) imaging as a testbed, we demonstrate 23× dynamic range extension and centimeter-level resolution from 3-bit quantized data, validated through simulations and hardware experiments. These results establish USF as an enabler of joint \emph{amplitude-time super-resolution}, opening new possibilities for low-resource sensing and imaging systems.

This work remains at an early stage, with several open challenges offering directions for future research. Theoretically, a rigorous analysis of \ussr under quantization would strengthen \thmref{thm:1} and provide practical guidelines, while a foundation for the empirical trends in \fig{fig:curve} would advance sparse super-resolution more broadly. Algorithmically, enhancing \sritersis by incorporating transient samples and sparse outliers \cite{Bhandari:2022:C}, together with a systematic study of its convergence, noise robustness, and performance bounds, would further support its applicability in sensing and imaging.

Taken together, the results and open directions highlight the potential of USF not only as a practical tool for next-generation acquisition systems but also as a foundation for new theoretical advances in sparse recovery and super-resolution.



\begin{thebibliography}{10}


\providecommand{\url}[1]{#1}
\csname url@samestyle\endcsname
\providecommand{\newblock}{\relax}
\providecommand{\bibinfo}[2]{#2}
\providecommand{\BIBentrySTDinterwordspacing}{\spaceskip=0pt\relax}
\providecommand{\BIBentryALTinterwordstretchfactor}{4}
\providecommand{\BIBentryALTinterwordspacing}{\spaceskip=\fontdimen2\font plus
\BIBentryALTinterwordstretchfactor\fontdimen3\font minus
  \fontdimen4\font\relax}
\providecommand{\BIBforeignlanguage}[2]{{%
\expandafter\ifx\csname l@#1\endcsname\relax
\typeout{** WARNING: IEEEtran.bst: No hyphenation pattern has been}%
\typeout{** loaded for the language `#1'. Using the pattern for}%
\typeout{** the default language instead.}%
\else
\language=\csname l@#1\endcsname
\fi
#2}}
\providecommand{\BIBdecl}{\relax}
\BIBdecl



\bibitem{Bhandari:2022:Book}
A.~Bhandari, A.~Kadambi, and R.~Raskar, \emph{Computational Imaging},
  1st~ed.\hskip 1em plus 0.5em minus 0.4em\relax MIT Press, Oct. 2022, open
  Access URL: https://imagingtext.github.io/.

\bibitem{DeFigueiredo:1982:J}
R.~J.~P. De~Figueiredo and C.-L. Hu, ``Waveform feature extraction based on
  {Tauberian} approximation,'' \emph{{IEEE} Trans. Pattern Anal. Mach.
  Intell.}, vol.~4, no.~2, pp. 105--116, Mar. 1982.

\bibitem{Li:2000:J}
L.~Li and T.~P. Speed, ``Parametric deconvolution of positive spike trains,''
  \emph{Ann. Statist.}, vol.~28, no.~5, Oct. 2000.

\bibitem{Seelamantula:2014:J}
C.~S. Seelamantula and S.~Mulleti, ``Super-resolution reconstruction in
  frequency-domain optical-coherence tomography using the
  finite-rate-of-innovation principle,'' \emph{{IEEE} Trans. Signal Process.},
  vol.~62, no.~19, pp. 5020--5029, Oct. 2014.

\bibitem{RedoSanchez:2016:J}
A.~Redo-Sanchez, B.~Heshmat, A.~Aghasi, S.~Naqvi, M.~Zhang, J.~Romberg, and
  R.~Raskar, ``Terahertz time-gated spectral imaging for content extraction
  through layered structures,'' \emph{Nature Communications}, vol.~7, no.~1,
  Sep. 2016.

\bibitem{Denoyelle:2019:J}
Q.~Denoyelle, V.~Duval, G.~Peyr{é}, and E.~Soubies, ``The sliding
  {Frank–Wolfe} algorithm and its application to super-resolution
  microscopy,'' \emph{Inverse Problems}, vol.~36, no.~1, p. 014001, Dec. 2019.

\bibitem{Pan:2017:J}
H.~Pan, T.~Blu, and M.~Vetterli, ``Towards generalized {FRI} sampling with an
  application to source resolution in radioastronomy,'' \emph{IEEE Trans.
  Signal Process.}, vol.~65, no.~4, pp. 821--835, Feb. 2017.

\bibitem{Batenkov:2020:J}
D.~Batenkov, G.~Goldman, and Y.~Yomdin, ``Super-resolution of near-colliding
  point sources,'' \emph{Information and Inference: A Journal of the IMA},
  vol.~10, no.~2, pp. 515--572, May 2020.

\bibitem{Vetterli:2002:J}
M.~Vetterli, P.~Marziliano, and T.~Blu, ``Sampling signals with finite rate of
  innovation,'' \emph{IEEE Trans. Signal Process.}, vol.~50, no.~6, pp.
  1417--1428, Jun. 2002.

\bibitem{Gedalyahu:2010:J}
K.~Gedalyahu and Y.~C. Eldar, ``Time-delay estimation from low-rate samples: A
  union of subspaces approach,'' \emph{IEEE Trans. Signal Process.}, vol.~58,
  no.~6, pp. 3017--3031, Jun. 2010.

\bibitem{Bredies:2012:J}
K.~Bredies and H.~K. Pikkarainen, ``Inverse problems in spaces of measures,''
  \emph{ESAIM: Control, Optimisation and Calculus of Variations}, vol.~19,
  no.~1, pp. 190--218, Mar. 2012.

\bibitem{Candes:2013:J}
E.~J. Candès and C.~Fernandez-Granda, ``Towards a mathematical theory of
  super-resolution,'' \emph{Comm. Pure Appl. Math.}, vol.~67, no.~6, pp.
  906--956, Apr. 2013.

\bibitem{Tang:2013:J}
G.~Tang, B.~N. Bhaskar, P.~Shah, and B.~Recht, ``Compressed sensing off the
  grid,'' \emph{IEEE Trans. Inf. Theory}, vol.~59, no.~11, pp. 7465--7490, Nov.
  2013.

\bibitem{Duval:2014:J}
V.~Duval and G.~Peyré, ``Exact support recovery for sparse spikes
  deconvolution,'' \emph{Foundations of Computational Mathematics}, vol.~15,
  no.~5, pp. 1315--1355, Oct. 2014.

\bibitem{Chi:2020:J}
Y.~Chi and M.~Ferreira Da~Costa, ``Harnessing sparsity over the continuum:
  {Atomic} norm minimization for superresolution,'' \emph{IEEE Signal Process.
  Mag.}, vol.~37, no.~2, pp. 39--57, Mar. 2020.

\bibitem{Weinstein:2013:J}
A.~J. Weinstein and M.~B. Wakin, ``Recovering a clipped signal in sparseland,''
  \emph{Sampling Theory in Signal and Image Processing}, vol.~12, no.~1, pp.
  55--69, Jan. 2013.

\bibitem{Walden:1999:J}
R.~Walden, ``Analog-to-digital converter survey and analysis,'' \emph{{IEEE} J.
  Sel. Areas Commun.}, vol.~17, no.~4, pp. 539--550, Apr. 1999.

\bibitem{Bhandari:2016:J}
A.~Bhandari and R.~Raskar, ``Signal processing for time-of-flight imaging
  sensors: An introduction to inverse problems in computational 3-d imaging,''
  \emph{{IEEE} Signal Process. Mag.}, vol.~33, no.~5, pp. 45--58, Sep. 2016.

\bibitem{Bhandari:2020:J}
A.~Bhandari, M.~H. Conde, and O.~Loffeld, ``One-bit time-resolved imaging,''
  \emph{{IEEE} Trans. Pattern Anal. Mach. Intell.}, vol.~42, no.~7, pp.
  1630--1641, Jul. 2020.

\bibitem{Guo:2025:J}
R.~Guo and A.~Bhandari, ``Blind {Time-of-Flight} imaging: Sparse deconvolution
  on the continuum with unknown kernels,'' \emph{SIAM Journal on Imaging
  Sciences}, vol.~18, no.~2, pp. 1439--1467, May 2025.

\bibitem{Bhandari:2017:Cb}
\BIBentryALTinterwordspacing
A.~Bhandari, F.~Krahmer, and R.~Raskar, ``On unlimited sampling,'' in
  \emph{Intl. Conf. on Sampling Theory and Applications ({SampTA})}, Jul. 2017.
\BIBentrySTDinterwordspacing

\bibitem{Bhandari:2020:Ja}
------, ``On unlimited sampling and reconstruction,'' \emph{{IEEE} Trans.
  Signal Process.}, vol.~69, pp. 3827--3839, Dec. 2020.

\bibitem{Bhandari:2021:J}
A.~Bhandari, F.~Krahmer, and T.~Poskitt, ``Unlimited sampling from theory to
  practice: {Fourier}-{Prony} recovery and prototype {ADC},'' \emph{{IEEE}
  Trans. Signal Process.}, pp. 1131--1141, Sep. 2021.

\bibitem{Bhandari:2020:C}
A.~Bhandari and F.~Krahmer, ``{HDR} imaging from quantization noise,'' in
  \emph{{IEEE} Intl. Conf. on Image Processing ({ICIP})}, Oct. 2020, pp.
  101--105.

\bibitem{Beckmann:2022:J}
M.~Beckmann, A.~Bhandari, and F.~Krahmer, ``The modulo {Radon} transform:
  Theory, algorithms, and applications,'' \emph{SIAM Journal on Imaging
  Sciences}, vol.~15, no.~2, pp. 455--490, Apr. 2022.

\bibitem{Beckmann:2024:J}
M.~Beckmann, A.~Bhandari, and M.~Iske, ``Fourier-domain inversion for the
  modulo {Radon} transform,'' \emph{IEEE Trans. Comput. Imaging}, vol.~10, pp.
  653--665, Apr. 2024.

\bibitem{Zhu:2024:Ca}
Y.~Zhu and A.~Bhandari, ``Unleashing dynamic range and resolution in unlimited
  sensing framework via novel hardware,'' in \emph{2024 IEEE SENSORS}.\hskip
  1em plus 0.5em minus 0.4em\relax IEEE, Oct. 2024, pp. 1--4.

\bibitem{Feuillen:2023:C}
T.~Feuillen, B.~S. M.~R. Rao, and A.~Bhandari, ``Unlimited sampling radar:
  {Life} below the quantization noise,'' in \emph{{IEEE} Intl. Conf. on
  Acoustics, Speech and Signal Processing (ICASSP)}, Jun. 2023.

\bibitem{Liu:2023:J}
Z.~Liu, A.~Bhandari, and B.~Clerckx, ``$\lambda$-{MIMO}: {Massive} {MIMO} via
  modulo sampling,'' \emph{{IEEE} Trans. Commun.}, pp. 6301 -- 6315, Nov. 2023.

\bibitem{Liu:2025:J}
------, ``Full-duplex beyond self-interference: {The} unlimited sensing way,''
  \emph{IEEE Communications Letters}, vol.~29, no.~1, pp. 165--169, Jan. 2025.

\bibitem{Guo:2024:J}
R.~Guo, Y.~Zhu, and A.~Bhandari, ``{Sub-Nyquist} {USF} spectral estimation:
  {$K$} frequencies with {$6K+4$} modulo samples,'' \emph{{IEEE} Trans. Signal
  Process.}, vol.~72, pp. 5065--5076, Sep. 2024.

\bibitem{Guo:2025:Cc}
R.~Guo and A.~Bhandari, ``{USF} spectral estimation: Prevalence of {Gaussian}
  {Cramér-Rao} bounds despite modulo folding,'' in \emph{IEEE Statistical
  Signal Processing Workshop (SSP)}, Jun. 2025, pp. 51--55.

\bibitem{Bhandari:2018:Ca}
A.~Bhandari, F.~Krahmer, and R.~Raskar, ``Unlimited sampling of sparse
  signals,'' in \emph{{IEEE} Intl. Conf. on Acoustics, Speech and Signal
  Processing (ICASSP)}, Apr. 2018.

\bibitem{Bhandari:2022:J}
A.~Bhandari, ``Back in the {US}-{SR}: {Unlimited} sampling and sparse
  super-resolution with its hardware validation,'' \emph{{IEEE} Signal Process.
  Lett.}, vol.~29, pp. 1047--1051, Mar. 2022.

\bibitem{Mulleti:2024:J}
S.~Mulleti and Y.~C. Eldar, ``Modulo sampling of {FRI} signals,'' \emph{IEEE
  Access}, vol.~12, pp. 60\,369--60\,384, Apr. 2024.

\bibitem{Itoh:1982:J}
K.~Itoh, ``Analysis of the phase unwrapping algorithm,'' \emph{Appl Optics},
  vol.~21, no.~14, p. 2470, Jul. 1982.

\bibitem{Guo:2023:Ca}
R.~Guo and A.~Bhandari, ``Unlimited sampling of {FRI} signals independent of
  sampling rate,'' in \emph{{IEEE} Intl. Conf. on Acoustics, Speech and Signal
  Processing (ICASSP)}, Jun. 2023.

\bibitem{Eamaz:2023:C}
A.~Eamaz, F.~Yeganegi, K.~V. Mishra, and M.~Soltanalian, ``Unlimited sampling
  of {FRI} signals with dithered one-bit quantization,'' in \emph{57th Asilomar
  Conference on Signals, Systems, and Computers}.\hskip 1em plus 0.5em minus
  0.4em\relax IEEE, Oct. 2023, pp. 1430--1435.

\bibitem{Kolmogorov:1949:J}
A.~Kolmogorov, ``On inequalities between the upper bounds of the successive
  derivatives of an arbitrary function on an infinite interval,''
  \emph{American Math. Soc. Trans.}, vol.~19, no.~4, 1949.

\bibitem{Deboor:1994:J}
C.~{De}~Boor, R.~Devore, and A.~Ron, ``The structure of finitely generated
  shift-invariant spaces in ${L_2(R^d)}$,'' \emph{Journal of Functional
  Analysis}, vol. 119, no.~1, pp. 37--78, Jan. 1994.

\bibitem{Aldroubi:1994:J}
A.~Aldroubi and M.~Unser, ``Sampling procedures in function spaces and
  asymptotic equivalence with {Shannon’s} sampling theory,'' \emph{Numerical
  Functional Analysis and Opt.}, vol.~15, no. 1–2, pp. 1--21, Jan. 1994.

\bibitem{Unser:2000:J}
M.~Unser and T.~Blu, ``Fractional splines and wavelets,'' \emph{SIAM Review},
  vol.~42, no.~1, pp. 43--67, Jan. 2000.

\bibitem{Nikolskii:1975:B}
\BIBentryALTinterwordspacing
S.~M. Nikol'ski\u{\i}, \emph{Approximation of Functions of Several Variables
  and Imbedding Theorems}.\hskip 1em plus 0.5em minus 0.4em\relax Springer
  Nature, 1975, vol. 205.
\BIBentrySTDinterwordspacing

\bibitem{Guo:2023:C}
R.~Guo and A.~Bhandari, ``{ITER-SIS}: {Robust} unlimited sampling via iterative
  signal sieving,'' in \emph{{IEEE} Intl. Conf. on Acoustics, Speech and Signal
  Processing (ICASSP)}, Jun. 2023.

\bibitem{Prony:1795:J}
G.~du~Prony, ``Essai experimental et analytique sur les lois de la dilatabilite
  de fluides elastiques et sur celles da la force expansion de la vapeur de
  l'alcool, a differentes temperatures,'' \emph{Journal de l'Ecole
  Polytechnique}, vol.~1, no.~22, pp. 24--76, 1795.

\bibitem{Steiglitz:1965:J}
K.~Steiglitz and L.~McBride, ``A technique for the identification of linear
  systems,'' \emph{{IEEE} Trans. Autom. Control}, vol.~10, no.~4, pp. 461--464,
  Oct. 1965.

\bibitem{Catala:2019:J}
P.~Catala, V.~Duval, and G.~Peyr\'{e}, ``A low-rank approach to off-the-grid
  sparse superresolution,'' \emph{SIAM Journal on Imaging Sciences}, vol.~12,
  no.~3, pp. 1464--1500, Jan. 2019.

\bibitem{Wang:2022:J}
W.~Wang, J.~Li, and H.~Ji, ``$l_1$-norm regularization for short-and-sparse
  blind deconvolution: Point source separability and region selection,''
  \emph{SIAM J. Imaging Sci.}, vol.~15, no.~3, pp. 1345--1372, Aug. 2022.

\bibitem{Blu:2008:J}
T.~Blu, P.-L. Dragotti, M.~Vetterli, P.~Marziliano, and L.~Coulot, ``Sparse
  sampling of signal innovations,'' \emph{{IEEE} Signal Process. Mag.},
  vol.~25, no.~2, pp. 31--40, Mar. 2008.

\bibitem{Kuo:2020:J}
H.-W. Kuo, Y.~Zhang, Y.~Lau, and J.~Wright, ``Geometry and symmetry in
  short-and-sparse deconvolution,'' \emph{SIAM J. Math. Data Sci.}, vol.~2,
  no.~1, pp. 216--245, Jan. 2020.

\bibitem{Perrone:2016:J}
D.~Perrone and P.~Favaro, ``A clearer picture of total variation blind
  deconvolution,'' \emph{{IEEE} Trans. Pattern Anal. Mach. Intell.}, vol.~38,
  no.~6, pp. 1041--1055, Jun. 2016.

\bibitem{Stoica:1989:J}
P.~Stoica and A.~Nehorai, ``{MUSIC}, maximum likelihood, and {Cramer-Rao}
  bound,'' \emph{{IEEE} Trans. Acoust., Speech, Signal Process.}, vol.~37,
  no.~5, pp. 720--741, May 1989.

\bibitem{Hua:1990:J}
Y.~Hua and T.~Sarkar, ``Matrix pencil method for estimating parameters of
  exponentially damped/undamped sinusoids in noise,'' \emph{{IEEE} Trans.
  Acoust., Speech, Signal Process.}, vol.~38, no.~5, pp. 814--824, May 1990.

\bibitem{Bhaskar:2013:J}
B.~N. Bhaskar, G.~Tang, and B.~Recht, ``Atomic norm denoising with applications
  to line spectral estimation,'' \emph{{IEEE} Trans. Signal Process.}, vol.~61,
  no.~23, pp. 5987--5999, Dec. 2013.

\bibitem{Park:2018:J}
Y.~Park, Y.~Choo, and W.~Seong, ``Multiple snapshot grid free compressive
  beamforming,'' \emph{The Journal of the Acoustical Society of America}, vol.
  143, no.~6, pp. 3849--3859, Jun. 2018.

\bibitem{Guo:2022:J}
R.~Guo, Y.~Li, T.~Blu, and H.~Zhao, ``Vector-{FRI} recovery of multi-sensor
  measurements,'' \emph{{IEEE} Trans. Signal Process.}, vol.~70, pp.
  4369--4380, Sep. 2022.

\bibitem{Pellegrini:2000:J}
S.~Pellegrini, G.~S. Buller, J.~M. Smith, A.~M. Wallace, and S.~Cova,
  ``Laser-based distance measurement using picosecond resolution
  time-correlated single-photon counting,'' \emph{Meas Sci Technol}, vol.~11,
  no.~6, pp. 712--716, May 2000.

\bibitem{Bhandari:2015:J}
A.~Bhandari, C.~Barsi, and R.~Raskar, ``Blind and reference-free fluorescence
  lifetime estimation via consumer time-of-flight sensors,'' \emph{Optica},
  vol.~2, no.~11, p. 965, Nov. 2015.

\bibitem{Heide:2013:J}
F.~Heide, M.~B. Hullin, J.~Gregson, and W.~Heidrich, ``Low-budget transient
  imaging using photonic mixer devices,'' \emph{ACM Trans. Graphics}, vol.~32,
  no.~4, pp. 1--10, Jul. 2013.

\bibitem{Shtendel:2022:C}
G.~Shtendel and A.~Bhandari, ``{HDR}-{ToF}: {HDR} time-of-flight imaging via
  modulo acquisition,'' in \emph{{IEEE} Intl. Conf. on Image Processing
  ({ICIP})}, Oct. 2022, pp. 3808--3812.

\bibitem{Kadambi:2013:J}
A.~Kadambi, R.~Whyte, A.~Bhandari, L.~Streeter, C.~Barsi, A.~Dorrington, and
  R.~Raskar, ``Coded time of flight cameras: sparse deconvolution to address
  multipath interference and recover time profiles,'' \emph{ACM Trans.
  Graphics}, vol.~32, no.~6, pp. 1--10, Nov. 2013.

\bibitem{Bhandari:2022:C}
A.~Bhandari, ``Unlimited sampling with sparse outliers: {Experiments} with
  impulsive and jump or reset noise,'' in \emph{{IEEE} Intl. Conf. on
  Acoustics, Speech and Signal Processing (ICASSP)}, May 2022, pp. 5403--5407.

\end{thebibliography}
\end{document}